# Programmes du FMI et performances économiques en RDC: Historique, impact et perspectives[*]

Matata Ponyo Mapon[†]

Jean-Paul K. Tsasa[‡]

**Résumé**

En fin 2012, la République démocratique du Congo (RDC) n'était plus en programme économique avec le Fonds monétaire international (FMI). Cependant, à la suite notamment des tensions inflationnistes survenues au dernier trimestre 2016, certains décideurs politiques avaient réclamé la reprise d'une coopération formelle avec le FMI. Cette démarche a finalement abouti en décembre 2019, lorsque le FMI a annoncé l'accord d'un programme de référence supervisé par son personnel. Cette annonce fut accueillie avec satisfaction par la classe politique dirigeante et abondamment commentée dans les médias. Pourtant, de la littérature économique, ressort-il que les effets des programmes du FMI sur la croissance économique sont généralement mitigés. Dans les rares cas où ces effets sont positifs, ils sont faibles ou statistiquement non significatifs. Ces évidences, apparemment contre-intuitives, peuvent davantage être renforcées dans le contexte de la RDC, où les meilleures performances économiques, depuis son indépendance en 1960, ont été réalisées entre 2012 et 2016; une période où le pays n'était justement pas en programme avec le FMI. D'aucuns se posent la question de savoir si l'assistance du FMI est-elle une malédiction pour les pays bénéficiaires? Nous soutenons que le problème de fond ne réside pas dans le recours ou dans la réfutation de l'assistance du FMI, mais plutôt dans la capacité des décideurs politiques à instaurer un leadership efficient et une gouvernance de qualité en vue de l'élaboration et de la mise en œuvre des réformes structurelles.

**Mots-clés**: FMI, RDC, Performances économiques, Leadership, Gouvernance.
**Classification JEL**: D78, E61, F33, O19.

**Abstract** (*IMF Programs and Economic Performance in the DRC: Documentation, Impact and Prospects*). At the end of 2012 the International Monetary Fund (IMF) has suspended its financial assistance to the Democratic Republic of the Congo (DRC). Due to inflationary pressures which occurred in the last quarter of 2016, several decision-makers called for a reopening of a formal cooperation with the IMF. This process was formally completed in December 2019. The restart of IMF programs was greeted with satisfaction by politicians and widely commented in the media. However, recent history shows that the DRC managed to achieve exceptional economic performance, between 2012 and 2016, without being in a formal cooperation with the IMF. Some people wonder whether IMF assistance is a curse for recipient countries? We argue that the underlying problem has nothing to do with accepting or not the IMF assistance, but rather in the ability of policy makers to establish effective leadership and good governance for the development and implementation supporting structural reforms.

**Keywords**: IMF, DRC, Economic performance, Leadership, Governance.
**JEL codes**: D78, E61, F33, O19.

*IMF loans react to economic conditions but are also sensitive to political-economy variables*. Barro and Lee (2005, p. 1245).

## 1. Introduction

En fin 2012, la République démocratique du Congo (RDC) n'était plus en programme économique formel avec le Fonds monétaire international (FMI). Cependant, à la suite notamment de la succession des gouvernements de cohabitation et de coalition depuis 2016[1], et des tensions inflationnistes qui s'en sont ensuivies, plusieurs décideurs politiques avait formellement formulé le vœu de relancer la coopération avec le FMI dans l'espoir de bénéficier d'une assistance à la fois technique et financière nécessaire pour préserver la stabilité macroéconomique et garantir un profil de croissance économique stable à moyen et long terme[2]. Cette démarche a finalement abouti en décembre 2019, avec l'accord d'un programme de référence (SMP) et l'approbation d'un décaissement de 368,4 millions de dollars américains au titre de la facilité de crédit rapide (FCR)[3], destinée à répondre aux besoins urgents de la balance des paiements[4].

---

[1] La succession des gouvernements de cohabitation et de coalition en RDC, durant la période 2016-2019, s'explique principalement par le report des élections présidentielles, législatives et provinciales initialement prévues en 2016 et des tensions qui s'en sont ensuivies. Le Gouvernement Matata, en place depuis 2012, a été successivement relayé par: (i) le Gouvernement Badibanga (19 décembre 2016-9 mai 2017) issu de l'accord politique de la cité de l'OUA; (ii) le Gouvernement Tshibala (9 mai 2017-6 septembre 2019) issu de l'accord politique global et inclusif (APGI), appelé également Accord de la Saint-Sylvestre. Les deux accords politiques, ont été respectivement signés le 18 octobre 2016 au siège de la cité de l'OUA sous la médiation de l'ex-premier ministre togolais Edem Kodjo, désigné « facilitateur » du dialogue par l'Union africaine (UA), et le 31décembre 2016 au Centre Interdiocésain de Kinshasa sous la médiation de la Conférence Épiscopale Nationale de la République démocratique du Congo (CENCO). Ces accords ont consacré le dialogue et la cohabitation entre la majorité présidentielle, et une frange de l'opposition et de la société civile jusqu'à la tenue des élections présidentielles, législatives et provinciales fixées en décembre 2018. Le Gouvernement Ilunga (6 septembre 2019-à ce jour) est issu de la coalition entre les forces politiques du candidat vainqueur de l'élection présidentielle du 30 décembre 2018 et celles de son prédécesseur qui ont préservé la majorité au parlement et dans les 26 provinces du pays (gouvernorat). Aussi, à souligner que, plus tôt, en date du 7 décembre 2014, le Gouvernement Matata avait connu un réaménagement technique à l'effet de mettre en exécution la principale conclusion du Comité de suivi des Concertations nationales, co-présidé par Aubin Minaku et Léon Kengo wa Dondo (respectivement présidents de la chambre basse et de la chambre haute du parlement en RDC), qui recommandait la formation d'un gouvernement de cohésion nationale dans l'objectif de planifier le développement socio-économique dans la paix et la concorde.

[2] Par exemple, en date du 12 juin 2017, le premier ministre Bruno Tshibala avait formellement sollicité une assistance du FMI dans le cadre de la FCR, en vue notamment d'atténuer l'accélération du niveau général des prix et de contenir le déficit public galopant (cf. Pilling 2017, Ross 2017).

[3] Il s'agit en effet d'un prêt à taux de 0% sur une durée maximale de 10 ans. Consécutivement à l'encaissement de cette FCR, les réserves internationales se sont établies à 1,03 milliard USD en fin décembre 2019 (soit 5 semaines d'importations des biens et services sur ressources propres), contre 660,03 millions USD en fin 2018 (cf. Banque centrale du Congo, 2020, p. 2). L'exécution du programme de référence (SMP, *staff monitored program*) est prévue pour la période de décembre 2019 à mai 2020.

[4] Voir FMI (2019c) et l'annexe C pour plus de détails.

L'annonce de la reprise du programme avec le FMI a été accueillie avec satisfaction par la classe politique dirigeante et, par ailleurs, largement commentée dans les médias locaux et étrangers. Pourtant, de la littérature économique, il ressort que les effets de la participation d'un pays aux programmes du FMI sur la croissance économique sont mixtes. Dans les rares cas où ces effets sont positifs, ils sont soit faibles, soit statistiquement non significatifs. Ces évidences, apparemment contre-intuitives, peuvent davantage être renforcées dans le contexte de la RDC. En effet, l'histoire récente montre que l'économie congolaise a réalisé des performances économiques exceptionnelles entre 2012 et 2016, une période où elle n'était justement pas en programme avec le FMI.

Tableau 1: Performances comparatives de l'économie congolaise avec et sans l'assistance du Fonds monétaires international

|  | 1980-1990 | 1991-2000 | 2001-2011 | 2012-2016 | 2017-2018 |
|---|---|---|---|---|---|
| Croissance en RDC nette de l'Afrique subsaharienne (%) | -0,44 | -7,71 | -0,93 | 2,60 | 1,63 |
| Inflation en RDC nette de l'Afrique subsaharienne (%) | 56,73 | 3891,62 | 18,20 | -1,04 | 32,64 |
| Élasticité-croissance du prix mondial du cuivre | . . . | 0,63 | 0,11 | -0,25 | 0,42 |
| Solde de la balance des paiements (en % PIB) | -0,52 | -6,95 | -2,89 | 0,11 | -1,40 |
| Assistance technique et financière du FMI | oui | non | oui | non | non |

Note: (i) La croissance en RDC nette de l'Afrique subsaharienne mesure la différence entre le taux de croissance du PIB par habitant en RDC et la moyenne du taux de croissance du PIB par habitant au niveau de l'Afrique subsaharienne. (ii) L'inflation en RDC nette de l'Afrique subsaharienne mesure la différence entre le taux d'inflation en RDC et la moyenne du taux d'inflation au niveau de l'Afrique subsaharienne. (iii) L'élasticité-croissance du prix mondiale du cuivre dénote l'élasticité du PIB par habitant en RDC par rapport au prix du cuivre sur le marché mondial des matières premières. Les détails explicatifs et autres indications sur les différentes sources des données sont repris en Annexe A (Voir Tableaux A.1, A.2 et A.3).

Comme nous l'illustrons dans le tableau 1[5], les performances économiques réalisées par la RDC entre 2012 et 2016 sont caractérisées de « exceptionnelles » pour au moins trois raisons. Premièrement, sur le plan régional, ces performances sont significativement supérieures à celles observées, en moyenne, dans l'ensemble des pays en Afrique subsaharienne. En effet, entre 2012 et 2016, le taux de croissance économique en RDC a été 2,60 points plus élevé que la moyenne du taux de croissance en Afrique subsaharienne. Par ailleurs, durant la même période, le taux d'inflation en RDC a été 1,04 point plus faible que la moyenne du taux d'inflation en Afrique subsaharienne[6]. Deuxièmement, sur le plan interne, la sous-période 2012-2016 demeure nettement distinctive comparativement aux autres sous-périodes. La tendance reste la même si nous prolongeons la période d'observation jusqu'en 1960. En effet, entre 1960 et 1979, la croissance et l'inflation en RDC nettes de l'Afrique subsaharienne ont été respectivement de -2,42% et 24,54%. Enfin, troisièmement, durant la période 2012-2016, les prix des matières premières, et notamment celui du cuivre, sur le marché mondial étaient en baisse monotone. Les mouvements du prix des matières premières, entre 2012 et 2016, ont eu une incidence négative, avec une élasticité-croissance du prix du cuivre se chiffrant à environ -0,25. Autrement dit, les performances économiques réalisées par la RDC durant cette période sont davantage remarquables car ne résultant pas du fait d'un environnement international clément[7]; mais découlant vraisemblablement d'un ensemble de politiques endogènes saines et minutieusement coordonnées[8], malgré le fait que le pays n'était pas en programme économique formel avec le FMI.

---

[5] Extrait du propos de David Lipton, Premier Directeur général adjoint du FMI: « J'ai été impressionné par les progrès accomplis ces cinq dernières années pour apporter la stabilité économique et une forte croissance, qui ont permis à la RDC d'enregistrer le troisième taux de croissance le plus rapide au monde en 2014 ». Cf. FMI (2015b).

Ainsi, le contraste apparent entre, d'une part, l'accueil satisfaisant de l'annonce de la reprise du programme avec le FMI par les acteurs politiques et les médias en décembre 2019 et, d'autre part, les évidences empiriques documentées dans la littérature économique, mais aussi la récente expérience de la RDC (2012-2016), nous pousse à adresser les questions que voici: (i) Au regard de l'attitude affichée par les acteurs politiques congolais, faut-il vraiment se réjouir de la reprise du programme avec le FMI? (ii) Au regard des évidences empiriques et de la récente expérience de la RDC, peut-on affirmer que l'assistance technique ou financière du FMI est, en règle générale, une nécessité absolue pour les pays en développement en proie aux problèmes d'instabilité macroéconomique?

Sur base des évidences empiriques tirées de l'expérience congolaise, nous montrons que le succès de la mise en œuvre des programmes du FMI dépend de manière critique, à la fois, de l'effectivité du leadership face aux groupes de pression et de qualité de la gouvernance dans la conduite des réformes transformationnelles. Plus important encore, nous soutenons qu'une fois que le leadership et la gouvernance de qualité garantissent l'effet de la reine rouge[9], les programmes du FMI cessent d'être une inhalation institutionnelle pour les pays en développement. À la lumière de ces évidences, nous discutons des politiques susceptibles de renforcer la solidité institutionnelle dans la conduite des réformes transformationnelles dans les pays en développement avec ou sans assistance des institutions de Bretton Woods. Par exemple, nous montrons que conditionnellement à un leadership éclairé et une gouvernance de qualité, les pays en développement, comme la RDC, n'auraient plus prioritairement besoin d'une assistance technique ou financière du FMI pour implémenter des réformes propices à la promotion d'une croissance inclusive ou à la réduction de l'incidence de la pauvreté; ils auraient surtout besoin des financements des projets spécifique de développement[10]. Malheureusement la « rigidité de règle » est telle qu'en général, pour les investisseurs et pourvoyeurs des fonds, le fait d'être en programme avec le FMI est un préalable au financement des projets de développement[11]. Ainsi, estimons-nous que le recours au programme du FMI doive être vu par les acteurs politiques non comme une fin en soi, mais fondamentalement comme un instrument qui permet de se conformer, le cas échéant, à cette « rigidité de règle » qu'un pays peut s'imposer avec ou sans le FMI.

---

[6] Les données utilisées proviennent de la base des indicateurs du développement dans le monde de la Banque mondiale.
[7] La RDC est un pays riche en ressources naturelles. Plusieurs auteurs tendent à expliquer sa dynamique économique par la cyclicité et la vigueur des prix des matières premières sur le marché mondial; voir par exemple: MacGaffey (1991), Akitoby et Cinyabuguma (2005), Dömeland et al. 2012.
[8] En effet, « politiques [...] minutieusement coordonnées » car entre 2012 et 2016, plusieurs plateformes d'interactions ont été mises en place, en plus des mécanismes classiques du travail interministériel que sont les commissions interministérielles permanentes. C'est le cas notamment de la Troïka stratégique (cf. Figure 3). Pour une discussion détaillée, voir Matata et Tsasa (2019, p. 28).
[9] Cf. Acemoglu et Robinson (2019, chap. 8).

En outre, il convient de souligner que le recours au programme du FMI n'est pas synonyme du « regain automatique » de la stabilité du cadre macroéconomique[12]. D'ailleurs, les

---

[10] En effet, contrairement aux banques de développement, le FMI n'accorde pas de prêts pour des projets spécifiques. Les concours financiers accordés par le FMI sont principalement destinés aux pays qui s'efforcent de: (i) reconstituer leurs réserves internationales; (ii) stabiliser la valeur de leur monnaie; (iii) continuer à régler leurs importations; (iv) restaurer les conditions d'une croissance forte et durable (FMI 2016b). Voir aussi Lacovara (1984) pour une discussion sur la convergence des rôles entre le FMI et la Banque mondiale (cf. Banque internationale pour la reconstruction et le développement, IBRD).

[11] À ce propos, de Looringhe et Ruben (2012, p. 221) notent: « ne pas accepter les changements apportés par le FMI veut dire pas d'accord dans le Club de Paris [...]. Cela compromettrait d'autres formes d'aide de la part des bailleurs de fonds qui exigent souvent un accord préalable avec le FMI ».

[12] À titre illustratif, supposons qu'un médecin généraliste prescrive une posologie préliminaire à un patient, par exemple, pour limiter les effets nuisibles d'une fièvre causée par une infection d'origine virale ou bactérienne. À ce stade, si le patient ne se conforme pas à la posologie, le médecin ne serait pas fautif en cas d'aggravation des symptômes. En sus, admettons que le médecin généraliste, en posant son premier diagnostic, recommande au patient de réaliser des examens approfondis pour identifier les causes du malaise et donc un traitement approprié. Une fois de plus, dans ce cas, le patient et son entourage direct sont les premiers responsables dans la réalisation de ces examens approfondis. Dans cette anecdote, le *médecin généraliste* correspond au FMI. Le *patient* c'est le pays bénéficiaire de l'assistance du FMI. Les *examens approfondis*, ce sont les diagnostics qui précèdent les réformes structurelles à mettre en œuvre. Tout comme les examens approfondis et le traitement y afférent sont non seulement dispendieux, mais aussi exigent de l'effort de la part du patient; de même, les coûts de la mise en œuvre des réformes structurelles sont généralement élevés et requièrent d'énormes sacrifices. Par ailleurs, il sied de remarquer que, dans cette histoire, le *respect de la posologie* est requis pour garantir un bon traitement de la maladie. De même, le succès du programme du FMI requiert suffisamment de discipline dans la mise en œuvre des politiques économiques et réformes structurelles. Finalement, s'il s'avérait que le patient considéré dans cette anecdote est très faible, il s'ensuit que le respect et le suivi de l'ensemble des procédures médicales que prescrites dépendront en grande partie de son *entourage direct*. De même, dans les pays en développement en proie à l'instabilité macroéconomique, les groupes de pression (tant sur le plan interne qu'externe) jouent un rôle majeur dans le succès ou l'échec des réformes structurelles mises en œuvre par les acteurs politiques. Que le pays soit en programme ou non avec le FMI, le rôle des groupes de pression ou lobbying reste important à cet effet. Ainsi, soutenons que le problème de fond dans le processus d'émergence des pays en développement ne réside pas dans le recours ou dans la réfutation de l'assistance du FMI ou d'autres bailleurs de fonds, mais réside plutôt, d'une part, dans l'effectivité du leadership des décideurs politiques (c'est-à-dire leur capacité à résister ou à négocier stratégiquement avec les différents groupes de

recherches disponibles à ce jour suggèrent que la contribution des programmes du FMI sur les performances économiques a été insignifiante, voire négative. Du point de vue théorique, ces évidences empiriques reflètent le résultat de la confrontation entre les effets positifs (*canaux directs*) et les effets négatifs (*canaux indirects*) que peuvent potentiellement exercer les programmes du FMI sur le profil de croissance des pays bénéficiaires d'une assistance technique ou financière.

En effet, d'une part, toute intervention du FMI devrait directement affecter le profil de croissance économique du pays bénéficiaire, pour au moins trois raisons (cf. Bird 1995, Killick 2003). Premièrement, le FMI donne des conseils en matière de politique en temps de crise. Suivre ces conseils devrait contribuer à améliorer la transparence, la gouvernance, le climat économique et ainsi favoriser la croissance future[13]. Deuxièmement, les prêts du FMI sont souvent assortis des conditions strictes[14], telles que la modification de l'exécution de la politique monétaire, l'élimination des contrôles de prix ou l'adoption d'une loi de finances conforme au cadre budgétaire du programme. D'ailleurs, en général, le décaissement effectif des prêts octroyés par le FMI n'a lieu que si le pays bénéficiaire respecte les conditionnalités sus-évoquées. De ce fait, accepter les conditionnalités imposées par le FMI devrait corollairement favoriser la croissance économique. Enfin, troisièmement, les fonds décaissés par le FMI en vue d'appuyer la balance des paiements devraient naturellement aider à assouplir les contraintes financières auxquelles sont confrontés les pays bénéficiaires et donc prévenir la stabilité du cadre macroéconomique.

D'autre part, toute assistance du FMI est susceptible d'affecter indirectement le profil de croissance économique du pays bénéficiaire, pour au moins deux raisons: du fait, d'une

---

pression en vue d'atteindre un objectif donné, dans ce cas un objectif d'intérêt général) et, d'autre part, dans la qualité de la gouvernance dans la conduite et le financement des réformes structurelles. Voir Eke et Kutan (2009) pour un exposé plus formel.

[13] En général, c'est dans le cadre des consultations au titre de l'article IV que le FMI s'emploie à donner des avis de politique générale sur les questions de gouvernance (FMI 2016e, 2017a). En sus, le FMI dispose de plusieurs mécanismes d'assistance technique (FMI 2016s, t, z, ad; 2018f, g), notamment: (i) l'instrument de soutien à la politique économique (ISPE, FMI 2017b), un mécanisme qui aide les pays à faible revenu à élaborer des programmes économiques efficaces; (ii) le mécanisme d'intégration commerciale (MIC, FMI 2016q), une politique conçue pour rendre plus prévisible l'accès aux financements du FMI dans le cadre des mécanismes de prêt existants; (iii) le suivi post-programme (FMI 2016r), un instrument permettant de déceler de manière rapide les politiques et contraintes qui pourraient compromettre la capacité d'un pays à rembourser la dette contractée auprès du FMI; (iv) le soutien aux pays à faible revenu (FMI 2016ae). Voir aussi FMI (2016u, v, w, x, y) pour plus de détails sur le programme d'assistance et d'évaluation du secteur financier en faveur des pays bénéficiaires.

[14] L'approbation d'un accord ou des revues d'un programme s'appuie sur divers engagements de politique économique convenus avec les autorités nationales (FMI 2016h, i; FMI 2018e). Il s'agit notamment des (i) mesures préalables, i.e. engagements à prendre avant l'approbation d'un financement par le Conseil d'administration du FMI ou l'achèvement d'une revue de programme; (ii) critères de réalisation quantitatifs, i.e. conditions spécifiques et mesurables se référant toujours à des variables macroéconomiques sur lesquelles le pays a une emprise; (iii) les objectifs indicatifs, pour évaluer quantitativement les progrès d'un pays membre par rapport aux objectifs d'un programme en cours; (iv) les repères structurels; i.e. des mesures de réforme (souvent non quantifiables) essentielles pour atteindre les objectifs du programme. Voir aussi Williamson (1983), Polak (1991), Guitian (1995), Eastonet Rockerbie (1999), Stiglitz (2001) et Dreher et Vaubel (2004).

part, de l'aléa moral (Vaubel 1983, Kapur 1998, Evrensel 2004, Lee et al. 2008) et, d'autre part, du syndrome hollandais (Paldam 1997, Doucouliagos et Paldam 2009). Suivant l'argument de l'aléa moral, le fait de pouvoir solliciter l'assistance du FMI s'apparente à une assurance. Cela peut inciter le pays bénéficiaire à adopter des politiques risquées ou peu judicieuses. Par ailleurs, l'hypothèse du syndrome hollandais suggère que les pays ayant d'importantes entrées de devises étrangères peuvent être soumis à une pression sur leur monnaie pour qu'elle s'apprécie, ce qui nuit à la compétitivité de leurs firmes sur les marchés internationaux. De ce qui précède, il ressort donc que si, empiriquement, la contribution des programmes du FMI sur les performances économiques dans les différents pays bénéficiaires s'est avérée négative ou non significative, c'est à cause notamment de la prédominance des canaux indirects sur les canaux directs.

Dans ce papier, nous soutenons que l'échec des programmes du FMI est fondamentalement causé par le déficit, à la fois, du leadership éclairé et de la gouvernance de qualité[15]. Nous définissons le leadership éclairé (ou efficace) comme étant la capacité d'un dirigeant politique (i) d'avoir une vision claire du progrès de son pays, partagée par l'ensemble ou la majeure partie de la population, (ii) de définir des objectifs globaux et intermédiaires précis et les mettre en œuvre, (iii) de résister aux pesanteurs de toute nature pouvant fragiliser ou anéantir la démarche entreprise et (iv) de vaincre les obstacles érigés par les groupes de pression et autres lobbyistes et rentiers qui cherchent à maintenir le pays dans une trappe compatible avec leurs intérêts privés; alors que l'objectif ultime pour un leader éclairé est d'améliorer de manière durable le bien-être collectif, i.e. de l'ensemble de la population. En parallèle, une gouvernance de qualité (ou simplement bonne gouvernance) peut être comprise comme étant la capacité du gouvernement à gérer efficacement ses ressources, à mettre en œuvre des politiques pertinentes et à promouvoir le respect des institutions par les citoyens et l'État, ainsi que l'existence d'un contrôle démocratique sur ceux qui gouvernent[16].

Le reste du papier s'organise comme suit. La section 2 propose un bref aperçu historique de la mise en œuvre des programmes économiques du FMI en RDC et discute sommairement des implications macroéconomiques des programmes du FMI dans les pays bénéficiaires tels que documentées dans la littérature économique. Plus spécifiquement les questions suivantes sont adressées: (i) Le progrès économique et social dans les pays en développement, et particulièrement en RDC, est-il vraiment lié à l'existence d'un programme avec le FMI ou non? Que nous renseignent les statistiques à cet effet? (ii) Le programme avec le FMI libère-t-il effectivement le potentiel économique des pays bénéficiaires ou l'étouffe plutôt? (iii) Peut-on être en programme avec les autres partenaires de développement, telles que la Banque mondiale ou la Banque africaines de développement, sans être en programme avec le FMI? Autrement dit, quid de la rigidité de règle? (iv) Les critères quantitatifs suffisent-ils pour garantir l'atteinte des objectifs assignés par le FMI aux pays bénéficiaires?

---

[15] Le binôme leadership éclairé-gouvernance de qualité est un concept difficile à caractériser et à quantifier. Dans Matata et Tsasa (2019), nous tentons de le quantifier en utilisant la base de données EPIN (évaluation de la qualité des politiques et des institutions nationales) du Groupe de la Banque mondiale (IDA 2017).
[16] Cf. Annexe F pour une discussion sur les définitions commodes, explicites et classiques du leadership et de la gouvernance. Ces définitions feront l'objet d'une analyse critique plus détaillée dans les recherches à venir.

La section 3 procède à une évaluation quantitative des corrélations contemporaine et retardées entre les programmes du FMI et les performances économiques en RDC, mesurées en termes de PIB per capita. Nous examinons également dans quelle mesure les différents programmes du FMI ont impacté le profil de croissance en RDC.

En référence aux récentes expériences de l'économie congolaise, la section 4 propose une brève discussion, d'une part, sur la dimension politique du FMI et, d'autre part, sur le rôle du leadership et de la gouvernance dans la conduite des reformes transformationnelles. La section 5 conclut.

## 2. Les programmes économiques du FMI en RDC: Historique et perspectives

Le FMI fut créé en juillet 1944, lors d'une conférence des Nations Unies à Bretton Woods dans le New Hampshire (États-Unis)[17], avec cinq objectifs spécifiques, à noter: (i) promouvoir la coopération monétaire internationale; (ii) garantir la stabilité financière; (iii) faciliter les échanges internationaux[18]; (iv) contribuer à un niveau élevé d'emploi et à la stabilité économique; (v) faire reculer la pauvreté (Clift 2004, p. 1)[19].

Au regard de ses objectifs, le FMI fournit une assistance technique ou financière à de nombreux pays développés et en développement[20], dont la RDC. Cette dernière a conclu son premier programme avec le FMI en juin 1967 pour une période de trois ans (juin 1967-juin 1970). Ce programme triennal consistait essentiellement à appuyer la réforme monétaire consacrant la nouvelle monnaie nationale, le zaïre, à la parité de 1 zaïre contre 1000 francs congolais et 2 dollars américains[21]. Le résultat de ce premier programme reste mitigé. Le taux de croissance du PIB par habitant et le taux d'inflation durant la période 1967-1970 se chiffrent à 0,22 et 33,34 pour cent respectivement contre une moyenne de 0,79 et 27,11 pour cent entre 1961 et 1966.

---

[17] La RDC y est membre depuis septembre 1963, avec une quote-part de 0,22%. La quote-part du Rwanda et celle des États-Unis sont de 0,03 et 17,46% respectivement. Pour plus de détails, voir FMI (2016c, 2016d, 2019b). L'annexe D explique brièvement comment fonctionne le FMI.

[18] Le FMI et l'Organisation mondiale du commerce (OMC) ont pour mission d'assurer la solidité du système de commerce et de paiements internationaux. Cependant, le FMI s'intéresse au système monétaire et financier international, alors que l'OMC se concentre sur le système mondial d'échanges commerciaux (cf. FMI 2016aa).

[19] En parallèle, dans un document publié par le Département de la communication du FMI en mars 2016, les objectifs du FMI se résument comme suit: (i) promouvoir la coopération monétaire internationale; (ii) faciliter l'expansion et la croissance équilibrées du commerce mondial; (iii) promouvoir la stabilité des changes; (iv) aider à établir un système multilatéral de paiements; (v) mettre ses ressources, moyennant des garanties adéquates, à la disposition des pays confrontés à des difficultés de balance des paiements (FMI 2016a, p. 3).

[20] Notons d'ores et déjà que le FMI et la Banque mondiale, tous deux institutions du système des Nations unies, poursuivent un même but: celui de relever le niveau de vie des populations des pays membres. Les approches utilisées à cet égard sont complémentaires: Alors que l'action du FMI est principalement centrée sur les questions macroéconomiques, la Banque mondiale se consacre au développement économique à long terme et à la lutte contre la pauvreté (cf. FMI 2016f, g; 2018b, c, d).

[21] Ministère des Affaires étrangères et de la Coopération internationale de la RDC (2007, pp. 4-5).

Par la suite, les chocs pétroliers et la baisse du cours du cuivre, durant la première moitié des années 1970, ont davantage affaibli les performances économiques de la RDC. Le taux de croissance est passé d'une moyenne positive de 0,96 pour cent durant la période 1961-1969 à une moyenne plus faible et négative de -0,70 pour cent entre 1970 et 1975. À l'effet de faire face à ces chocs, la RDC a pour une deuxième fois sollicité l'appui du FMI. Ainsi les deux partis ont formellement conclu, en mars 1976, un programme de stabilisation d'un an, appuyé par un accord de confirmation. Ce programme avait pour but d'aider le pays à préserver les grands équilibres macroéconomiques. Malheureusement, il fut suspendu à cause du non-respect des critères liés à la baisse des dépenses publiques et à l'obligation de couvrir le service de la dette.

Figure 1: Programmes du FMI et performances économiques en RDC

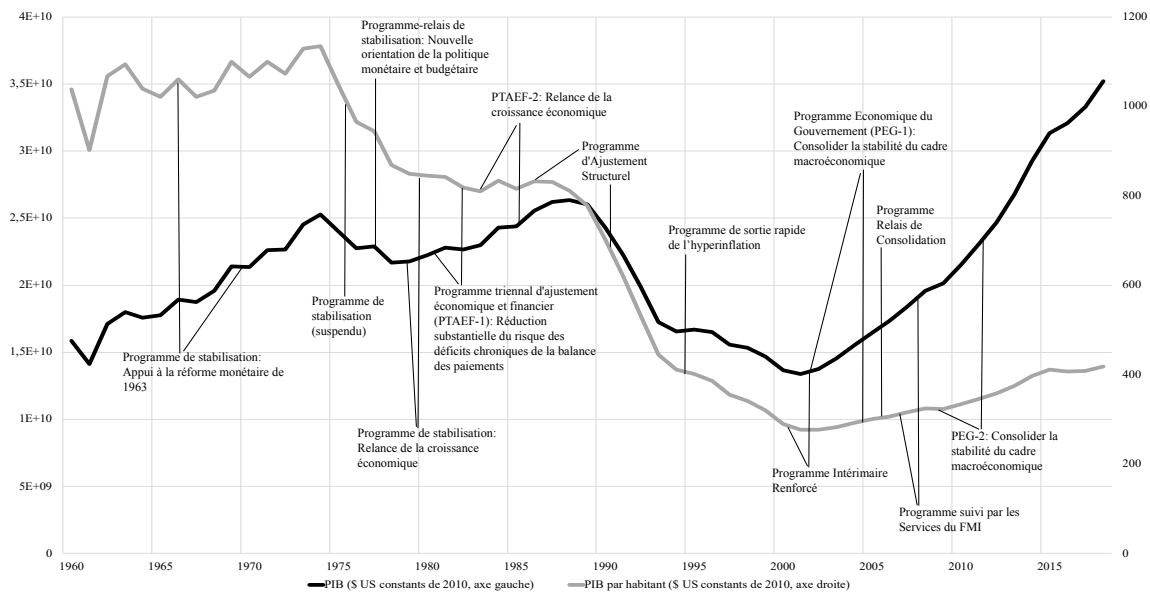

Note: Les données utilisées pour le PIB et le PIB par habitant proviennent de la base de données de la Banque mondiale sur les indicateurs du développement dans le monde (WDI DataBank 2019). Le PIB et le PIB par habitant sont mesurés en dollar américain constant de 2010. Les informations sur les différents programmes du FMI en RDC, entre 1963 et 2019, ont été recueillies dans les documents ci-après élaborés en vertu de l'article IV des statuts du FMI: Communiqué de presse; Rapport des services du FMI; Annexe d'informations des services du FMI; Déclaration du Directeur exécutif pour la RDC; Note d'information au public (FMI 2019a).

Un troisième programme formel, relayant le deuxième, fut conclu en avril 1977 pour une durée d'une année[22]. Également appuyé par un accord de confirmation, ce programme avait pour objectifs de soutenir la relance économique, notamment en procédant au passage du régime de change fixe vers un régime de change flexible et en proposant une nouvelle orientation des politiques budgétaires et monétaires. Une fois de plus, les résultats de ce programme, en termes de croissance économique et de stabilité du niveau général des prix restent mitigés. En effet, entre 1977 et 1979, les moyennes du taux de croissance du PIB par habitant et du taux d'inflation se chiffrent à -4,16 et 55,82 pour cent contre -7,95 et 57,40 pour cent en 1976.

---

[22] Cf. Banque mondiale (1979, p. 2).

Tableau 2: Aperçu des différents programmes économiques formels entre le FMI et la RDC (1960-2020)

| Programme avec le FMI | Début | Fin | Durée d'exécution (Nombre d'années) | Croissance du PIB par hab. (%) |
|---|---|---|---|---|
| 1. Programme triennal d'assistance financière et technique | 1967 | 1970 | 3 | 0,22 |
| 2. Programme de stabilisation et Accord de confirmation | 1976 | 1976 | Suspendu. Prévu pour 1 an | -7,95 |
| 3. Programme de stabilisation et Accord de confirmation | 1977 | 1977 | Suspendu. Prévu pour 1 an | -2,10 |
| 4. Programme de stabilisation et Accord de Stand-by | 1979 | 1980 | 1 | -1,42 |
| 5. Programme triennal d'ajustement économique et financier | 1981 | 1983 | 3 | -1,39 |
| 6. Programme triennal d'ajustement économique et financier | 1984 | 1986 | 3 | 0,94 |
| 7. Programme triennal d'ajustement économique et financier | 1987 | 1989 | 3 | -2,24 |
| 8. Programme triennal d'ajustement économique et financier | 1990 | 1992 | Non exécuté. Suspendu | -11,77 |
| 9. Programme de sortie rapide de l'hyperinflation | 1995 | 1996 | 1 | -3,13 |
| 10. Programme intérimaire suivi par les services du FMI (programme de référence) | 2001 | 2002 | 1 | -2,40 |
| 11. Programme économique du gouvernement soutenu par la FRPC | 2002 | 2005 | Suspendu en 2005 | 2,15 |
| 12. Programme relais de consolidation | 2006 | 2006 | 2/3 | 1,98 |
| 13. Programme économique suivi par les services du FMI | 2007 | 2008 | 1 | 2,82 |
| 14. Programme économique du gouvernement | 2009 | 2012 | Suspendu en fin 2012 | 2,51 |
| 15. Programme de référence | 2019 | 2020 | 1/2 | 1,10 |

En 1979, un quatrième programme de stabilisation (juillet 1979-décembre 1980) fut signé. Ce programme avait pour but d'assainir la situation financière du pays afin de rétablir la solvabilité du pays et de promouvoir la reprise de la croissance économique[23]. Ce programme fut également un échec apparent: le taux de croissance du PIB par habitant est demeuré négatif (-1,42 pour cent) et le taux d'inflation excessivement élevé (supérieur à 75 pour cent, hyperinflation).

En outre, comme nous pouvons le voir dans la figure 1, avant la suspension de toute coopération formelle entre la RDC et le FMI à la fin des années 1980, trois autres programmes furent formellement conclus, respectivement en 1981, 1984 et 1987. En 1981, un accord d'assistance technique et financière entre la RDC et le FMI avait été signé, dénommé programme triennal d'ajustement économique et financier (PTAEF-1, 1981-1983). L'objectif du PTAEF-1 était de relancer la croissance économique, de contenir la pression inflationniste et de réduire substantiellement le risque des déficits chroniques de la balance des paiements. Une fois de plus, l'application de ce programme n'a pas produit les résultats escomptés.

À la suite du PTAEF-1, la RDC avait sollicité un autre programme d'ajustement économique et financier, également d'une durée de trois ans (PTAEF-2, 1984-1986), afin, d'une part, de résoudre le problème de déficit budgétaire et de la balance des paiements et, d'autre part, de relancer la croissance économique et d'honorer le service de la dette. En 1987, le PTAEF-2 fut relayé par le programme d'ajustement structurel dans le but de réduire les déséquilibres budgétaires à court et à moyen termes et d'adapter la structure de l'économie nationale aux exigences d'une croissance stable à long terme[24]. Il était prévu que ce programme soit exécuté en deux séquences, entre 1987 et 1990 (*séquence 1*), puis entre 1990 et 1992 (*séquence 2*). Ce programme, partiellement exécuté, fut un échec. Cette fois-ci, le non-respect des conditionnalités imposées par le FMI conduisit à l'interruption de toute coopération formelle entre la RDC et le FMI durant la période 1991-2000. Toutefois en 1995, un programme de sortie rapide de l'hyperinflation fut exécuté avec le concours technique du FMI. Ce programme avait pour objectif de réduire substantiellement le niveau du taux d'inflation, de contenir les dérapages des finances publiques et de renouer avec une croissance positive[25]. Le programme de sortie rapide de l'hyperinflation ne fut malheureusement pas relayé par une assistance subséquente du FMI à cause de nombreuses contraintes institutionnelles causées par l'éclatement de la première guerre du Congo (1996-1997).

Après avoir été suspendue pour une période de près de dix ans (1991-2000), la coopération formelle entre la RDC et le FMI ne fut effectivement reprise qu'au début du second trimestre de 2001, avec la mise en œuvre successive de: Programme intérimaire renforcé (PIR) destiné à casser l'hyperinflation et à libéraliser l'économie (mai 2001-mars 2002); Programme économique du gouvernement (PEG-1, avril 2002-juillet 2005) destiné à consolider la stabilité du cadre macroéconomique, à promouvoir une croissance économique soutenable et à soutenir les efforts de réduction de la pauvreté. Le PEG-1 fut malheureusement arrêté sans avoir conclu les cinquième et sixième revues; Programme relais de consolidation (PRC, avril-décembre 2006), appuyé par un programme économique suivi par les services du FMI (PSSF) dont l'exécution s'est révélée peu satisfaisante à cause essentiellement du contexte conduisant aux élections générales dans le pays[26]; Programme économique suivi par les services du FMI (PSSF, 2007-2008); Programme économique du gouvernement (PEG-2, 2009-2012)[27]. Bien que tous les critères quantitatifs fussent satisfaits dans le cadre du PEG-2[28], la coopération formelle et effective entre la RDC et le FMI fut à nouveau interrompue en décembre 2012 à la suite, principalement, d'une divergence dans l'interprétation d'un critère qualitatif se rapportant à la transparence dans la gestion des ressources naturelles[29].

---

[23] Cf. Banque mondiale (1981, p. 27).
[24] Notamment la discipline budgétaire, la privatisation des entreprises et établissements publics, la libéralisation des prix et la réduction des barrières commerciales.
[25] Beaugrand (1997, p. 23).

Cette interruption durera près de sept ans, avant de reprendre en 2019. En effet, à l'issue des élections du 30 décembre 2018, le souhait de reprendre la coopération avec le FMI s'était davantage manifesté et, finalement, concrétisé à la suite d'une rencontre entre le Président de la RDC Félix-Antoine Tshisekedi et la Directrice générale du FMI Christine Lagarde, le 4 avril 2019, aux États-Unis. Après cette rencontre, madame Lagarde a déclaré: « Je suis ravie qu'on ait pu renouer la relation et mettre en place ensemble ce partenariat pour travailler à l'amélioration de la situation économique et celle de la population ». Le 26 août 2019, le Conseil d'administration du Fonds a conclu les consultations au titre de l'article IV avec la République démocratique du Congo (RDC). À la suite de la transition politique pacifique qui a eu lieu en début 2019, la reprise de la coopération entre la RDC et le FMI est perçue, par plusieurs observateurs et analystes[30], comme une occasion de mettre en place des réformes propices à la création d'emplois, la promotion d'une croissance inclusive et la réduction de l'incidence de la pauvreté.

De manière générale et comme il en ressort de la figure 2, l'incidence des programmes du FMI sur les performances économiques semble remarquablement mitigée. Par exemple durant la première vague des programmes avec le FMI (1967-1991), la moyenne du taux de croissance du PIB par habitant est négative (-2,03 pour cent). Ce taux est positif durant la deuxième vague (2001-2012, une moyenne de 1,79 pour cent), mais demeure largement inférieur à la moyenne du taux de croissance durant la période 2012-2016 où toute coopération formelle entre la RDC et le FMI était suspendue. En outre il convient également de noter que le fait d'être ou non en programme avec le FMI ne suffit pas à expliquer de manière cohérente les performances observées en RDC. Par exemple, entre 1991 et 2000, la RDC n'était pas en programme avec le FMI. Le taux de croissance du PIB par habitant durant cette période, i.e. 1991-2000, était de -7.89 pour cent, alors qu'entre 2012-2016, ce taux se chiffrait à 3,39 pour cent.

---

[26] Cf. de Looringhe et Ruben (2012, p. 109).

[27] Janvier-juin 2009: Programme économique suivi par les services du FMI (programme de référence). Pour plus de détails, cf. Lettre d'intention du 30 novembre 2009 adressée par le premier ministre Muzito à Dominique Strauss-Kahn, directeur général du FMI.

[28] Le PEG-2 avait expiré le 10 décembre 2012 (FMI 2014, p. 5). Voir FMI (2009a, pp. 26-34) pour un aperçu des critères de réalisation quantitatifs tels que présentés dans le mémorandum technique d'exécution du programme soumis par le gouvernement congolais au FMI, le 30 novembre 2009.

[29] Approuvé par le Conseil d'administration du FMI le 11 décembre 2009, le programme économique du Gouvernement (2009-2012) visait à aider le pays à surmonter les obstacles identifiés dans la Stratégie pour la réduction de la pauvreté et pour la croissance (cf. FMI 2009a, p. 2; BAD 2010, p. 5). Se rapporter à FMI (2009b) et Beith et Richardson (2015) pour plus de détails.

[30] Voir, par exemple, Ames et al. (2019), Lagarde (2019), William (2019a, b).

Figure 2: Performances économiques en RDC avec ou sans assistance du FMI

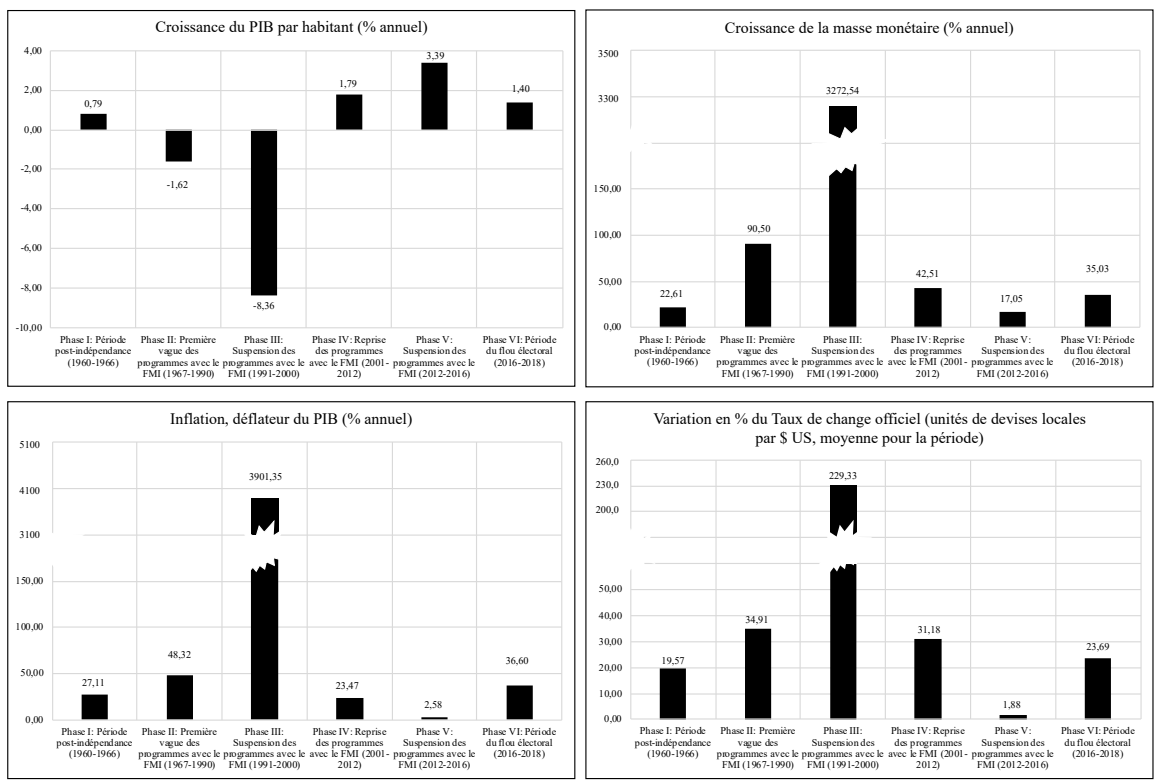

Note: Les données utilisées pour élaborer les quatre graphiques constitutifs de la figure 2 proviennent de la base des Indicateurs du développement dans le monde de la Banque mondiale (WDI DataBank 2019). Les échelles dans les axes des ordonnées pour les graphiques 2, 3 et 4 ont été adaptées pour, d'une part, prendre en compte les valeurs élevées qu'affichent les différents agrégats retenus dans la phase III (1992-2001) et, d'autre part, ne pas étouffer la lecture visuelle du niveau de ces mêmes agrégats durant les autres phases.

En outre, l'analyse historique du cas de la RDC ne fournit aucune preuve que le fait d'être en programme avec le FMI libère le potentiel économique du pays bénéficiaire. Enfin, à cause de la rigidité de règle[31], il est quasiment impossible pour les pays en développement de bénéficier de l'appui financier ou logistique conséquent de la part des autres partenaires de développement, telles que la Banque mondiale, la Banque africaine de développement, la Commission européenne ou des donateurs bilatéraux, sans préalablement être en programme avec le FMI. Toutefois, c'était exceptionnellement le cas de la RDC entre 2012 et 2016, au regard notamment des performances macroéconomiques réalisées par le pays durant cette période[32].

---

[31] Par rigidité de règle, nous faisons allusion au principe d'après lequel: Être en programme économique formel avec le FMI constitue, pour les institutions financières internationales traditionnelles telles que la Banque mondiale ou la Banque africaine de développement, une sorte de caution permettant d'accéder facilement aux financements.

Le caractère controversé des programmes du FMI n'est pas le propre de la RDC (Doug et Visquez 1994, Danaher 1995). Comme nous l'avons mentionné à l'introduction, ces controverses ont également été documentées et largement débattues dans la littérature économique. Par exemple, Przeworski et Vreeland (2000, p. 403) établissent que les pays qui n'adhèrent pas aux programmes du FMI croissent plus vite que ceux qui y participent, même lorsque les deux groupes sont confrontés à des conditions initiales similaires. Vreeland (2003) trouve que la croissance économique est environ 1,5% plus lente lorsque les pays bénéficient d'un programme du FMI[33]. D'où la question, pourquoi les gouvernements et le FMI continuent-ils de conclure des programmes d'assistance d'autant plus que le résultat est souvent décevant sur le plan économique. L'auteur estime que les gouvernements, en concluant des accords avec le FMI, recherchent avant tout un soutien politique afin de faire face notamment à l'opposition nationale[34]. Le même argument est soutenu par Pastor (1987, 2019).

Tableau 3: Effet des programmes du FMI sur la croissance économique

| Papier/Méthodologie | Négatif | Aucun | Positif |
| --- | --- | --- | --- |
| Before-After (Avant/Après) | 0 | 6 | 3 |
| With-Without (Avec/sans) | 1 | 6 | 1 |
| Régression | 7 | 5 | 3 |

Note: Tableau élaboré sur base des informations recueillies dans le tableau E.1 (cf. Annexe E).
Source: Dreher (2004, p. 31), Dreher (2006, p. 773).

Quels que soient l'approche méthodologique considérée, la période d'intérêt ou l'échantillon des pays, les résultats établis par Przeworski et Vreeland (2000) et la littérature subséquente semblent robustes. En instrumentant les variables de prêt du FMI avec les valeurs (i) de prêt prévu en fonction des quotas et des parts du personnel du FMI, (ii) des fractions de votes des Nations Unies avec les États-Unis et les pays européens, et

---

[32] En effet, comme nous l'avons expliqué dans la section introductive, la règle d'or (rigidité de règle) implique qu'un pays ne peut être en programme avec les autres institutions financières et partenaires techniques que s'il est en programme avec le FMI, considéré comme le visa d'accès aux sources de financement. Pour les bailleurs et pourvoyeurs de fonds, la philosophie de ce visa est de s'assurer d'une bonne gestion des projets de développement à financer. Toutefois, lorsque les partenaires de développement remarquent que le pays gère rigoureusement son économie, ce qui fut notamment le cas entre 2012 et 2016 en RDC, l'accès au financement peut donc être accordé.
[33] Vreeland (2003), cité par Stiles (2004, p. 644).
[34] Hülsmann (2003, p. 54) note par exemple: « on peut très bien se passer du FMI. Sa suppression immédiate serait certes déplorable du point de vue des classes politiques, mais bénéficierait à la large majorité des populations en Occident et en Afrique ».

(iii) de l'intensité des échanges avec les États-Unis et les pays européens, Barro et Lee (2005) corroborent les conclusions de Przeworski et Vreeland (2000). En utilisant des données de panel pour 98 pays sur la période 1970-2000, Dreher (2004, 2006) trouve des preuves que la participation aux programmes du FMI réduit la croissance économique. En outre, l'auteur établit que rien ne prouve que le respect de la conditionnalité atténuerait cet effet négatif. Par ailleurs, Dreher et Gassebner (2012, p. 330) constatent que les gouvernements sont confrontés à un risque croissant d'entrer en crise lorsqu'ils restent sous le régime du FMI une fois que les performances économiques s'améliorent. Les auteurs notent que ce constat peut s'expliquer par le fait que, d'une part, l'assistance technique ou financière dénote le symptôme de l'incompétence du gouvernement et, d'autre part, les gouvernements qui héritent des programmes du FMI pourraient être moins susceptibles de mettre en œuvre les conditionnalités convenues par leurs prédécesseurs.

Sur base des statistiques informatives, Easterly (2005, p. 20) conclut que l'ajustement structurel n'a pas réussi à ajuster la politique macroéconomique et à générer une croissance économique stable dans les pays bénéficiaires. En utilisant les données de 104 accords du FMI dans 74 pays en développement entre 1973 et 1991, Santaella (1996, p. 539) trouvent que les pays qui sont en programme ou sur le point de suivre un programme avec le FMI affichent une balance des paiements, une croissance économique, des conditions extérieures plus fragiles que ceux qui ne sont en programme[35]. Feldstein (1998) formule trois critiques à l'endroit des remèdes qu'a proposés le FMI à la crise asiatique, lesquelles peuvent toutefois être généralisées dans le cas des pays en développement. Premièrement, Feldstein note que le FMI prescrit généralement et parfois de manière déguisée la même vieille recette d'austérité[36], souvent de manière inappropriée aux pays souffrant des maladies diverses et singulières. Deuxièmement, il soutient qu'en incluant dans son programme un certain nombre d'éléments structurels, le FMI va généralement au-delà de sa tâche essentielle, à noter celle de corriger la balance des paiements, en s'immisçant dans les processus politiques des pays bénéficiaires de son assistance[37]. Troisièmement, il évoque le problème de l'aléa moral qu'impliquent généralement les programmes du FMI. Dans le même sens, Meltzer (1999) note que le FMI a connu plus de succès dans les années 60, lorsqu'il a limité ses efforts à l'aide aux pays souffrant de déficits courants. À mesure que la portée de ses interventions s'est élargie, son bilan est devenu ambigu.

---

[35] Ces résultats semblent intuitifs car les prêts du FMI sont accordés en réponse à une crise économique cf. Abbotta et al. (2010, p. 19).

[36] Alors qu'abordant dans le même sens que Feldstein (1998), Marysse (2010, p. 150) évoque l'effet Matthieu, Bird et Rowlands (2017, p. 2192) réfutent, en revanche, l'affirmation selon laquelle les programmes concessionnels du FMI dans les pays à faible revenu sont fondés sur l'austérité, même s'ils constatent une différence importante entre les programmes concessionnels et non concessionnels.

[37] En réponse à cette critique, Fischer (1998, p. 103) estime qu'éviter les aspects structurels et politiques dans le programme du FMI serait synonyme d'un traitement superficiel du problème de fond caractérisant les économies des pays qui sollicitent l'assistance du FMI et donc impliquerait vraisemblablement une récurrence chronique des crises. Autrement dit, les critères quantitatifs ne suffisent pas de garantir l'atteinte des objectifs assignés par le FMI aux pays bénéficiaires. Par ailleurs, à ce jour, certains pays émergents identifient des mécanismes d'emprunt alternatifs qui ne dépendent pas des réformes politiques contraignantes telles que recommandées par le FMI (Dreher et al. 2008, p. 167). Pour une revue des critiques traditionnellement adressées aux différents programmes économiques exécutés par le FMI dans les pays en développement, voir Helleiner (1983), Zaki (2001), Nooruddin et Simmons (2006), Conway (2007) et Pop-Eleches (2009).

Pour vérifier la robustesse des inférences concernant l'efficacité des programmes économiques du FMI, Eke et Kutan (2009) appliquent plusieurs méthodes d'évaluation[38], couramment utilisées dans la littérature, à un panel plus large des pays, incluant notamment les pays de l'Europe de l'Est. Les auteurs trouvent que les programmes du FMI dans les pays bénéficiaires sont inefficaces. Cependant, précisent-ils, qu'il peut être simpliste de blâmer uniquement le FMI pour l'échec de ces programmes car, en réalité, l'inefficacité de ces programmes peut être causée par de nombreux facteurs, notamment la nature des programmes, l'engagement des fonctionnaires ou du gouvernement du pays bénéficiaire, les changements des fondamentaux politiques ou encore les chocs externes. En utilisant un ensemble de données contenant des informations sur 94 programmes du FMI entre 1989 et 2002, Baqir et al. (2005) constatent que les implications macroéconomiques des assistances techniques et financières du FMI sont généralement inférieures aux attentes en matière de croissance et d'inflation.

En générale, les preuves empiriques présentées dans ce papier ne sont pas favorables au FMI[39]. En même temps, nous sommes d'accord notamment avec Dicks-Mireaux et al. (2000) et Eke et Kutan (2009) que l'immense complexité qu'implique les différentes approches méthodologiques et la nature des données statistiques utilisées dans la littérature, à ce jour, ne permet pas aux économistes de fournir une base solide pour tirer des conclusions sans ambiguïté quant aux implications macroéconomiques des programmes d'assistance technique et financière du FMI. Dans ce papier, nous tentons d'examiner l'impact des programmes du FMI sur la croissance économique en RDC, en tenant compte du biais d'endogénéité et du biais d'instantanéité. Nous utilisons les données de la banque mondiale et les informations fournies par le FMI pour permettre une réplication facile de l'ensemble de nos résultats.

## 3. Implications macroéconomiques des programmes du FMI en RDC

L'analyse empirique de l'interaction entre l'assistance du FMI et la croissance économique requiert beaucoup de prudence. Dans ce papier, nous accordons une attention particulière au biais d'endogénéité et au biais d'instantanéité. En effet, les pays qui sollicitent l'assistance du FMI sont généralement ceux qui font déjà face à des difficultés économiques imminentes au moment où ils soumettent leur demande[40]. De même, en

---

[38] Dans leur papier, les auteurs utilisent respectivement: (i) les deux méthodes alternatives que proposées par Bird (2001), à noter *Recidivism and Program Completion*; (ii) l'analyse *Before-and-After*; (iii) la méthode d'évaluation généralisée (cf. Eke et Kutan 2009, p. 10).

[39] En utilisant un ensemble de données incluant 213 pays, pour la période 1971-2009, Fidrmuc et Kostagianni (2015) trouvent que l'effet contemporain de la participation du FMI est non significatif tandis que l'effet décalé est positif. Le tableau E.1, en Annexe E (voir aussi le tableau 3), présente un échantillon de 32 études sur les effets des programmes du FMI sur la croissance. Sur les 32 études : 25 trouvent un effet négatif ou statistiquement non significatif et 7 un effet positif.

[40] Que leurs origines soient intérieures ou extérieures, les difficultés économiques ou crises peuvent revêtir des formes diverses: (i) Difficultés de balance des paiements, se produisant lorsqu'un pays n'est plus en mesure de régler ses importations essentielles ou ne peut plus assurer le remboursement de sa

général, parmi les pays qui demandent une assistance financière du FMI, et particulièrement ceux qui reçoivent un soutien, ont tendance à être dans une situation économique pire que ceux qui ne le font pas. Dès lors, la relation non significative, voire négative, entre l'assistance du FMI et la croissance économique, telle que documentée dans la littérature, peut être due à un tel biais d'endogénéité.

De ce qui précède, nous utilisons des variables instrumentales afin de tenir compte de l'endogénéité. Comme dans Drehel et al. (2009), nous nous concentrons sur des instruments qui reflètent les conditions institutionnelles ou politiques plutôt qu'économiques. Théoriquement, les instruments doivent être non corrélés avec le terme d'erreur. Par ailleurs, les instruments doivent posséder un pouvoir explicatif suffisant pour expliquer l'assistance du FMI sans être eux-mêmes corrélés à la croissance afin de permettre à l'analyste de les exclure de l'équation de régression principale. Nous utiliserons, à cet effet, les subventions de coopération technique (SCT), lesquelles sont indépendantes d'un prêt et ont pour but de financer le transfert de compétences techniques et en gestion ou de technologies afin de renforcer la capacité nationale générale sans référence à un projet spécifique d'investissement, mais aussi la capacité à exécuter des projets d'investissement spécifiques. Le deuxième instrument que nous utilisons, ce sont les subventions, hors coopération technique (SHCT). Les données pour ces deux instruments sont disponibles pour la période 1960-2018.

Tableau 4: Correspondance entre les variables du modèle et leurs équivalents sur données

| Variable | | Définition | Source |
|---|---|---|---|
| $Y/L$ | : | PIB réel divisé par la population en âge de travailler | Banque mondiale (2019) |
| $n$ | : | Taux de croissance de la population en âge de travailler | Banque mondiale (2019) |
| $s_k$ | : | Part de l'investissement (incluant l'investissement du gouvernement) dans le PIB réel | Banque mondiale (2019) |
| $\Phi_{FMI}$ | : | Participation à un programme du FMI | Tableau 2 (cf. *Section 2*) |

Par ailleurs, si l'assistance du FMI venait à favoriser la croissance, il est vraisemblable que cet effet positif ne se manifesterait qu'avec un certain retard (cf. Killick et al. 2012,

---

dette; (ii) Crises financières, surgissant lorsque les établissements financiers sont insolvables ou manquent de liquidités; (iii) Crises budgétaires, survenant lorsque le déficit budgétaire et l'endettement sont excessifs. Les crises provoquent souvent une hausse du chômage, une baisse des revenus et une augmentation de l'incertitude. Cf. Fmi (2015a) pour plus de détails sur les facteurs intérieurs et extérieurs à l'origine des crises et sur comment les financements du FMI peuvent donner une marge de manœuvre suffisante, aux pays en difficulté, pour les ajustements nécessaires.

Clemens et al. 2012). En effet, l'impact du programme du FMI peut être retardé pour au moins deux raisons: (i) certains programmes peuvent être effectivement exécutés tard dans l'année et donc ne peuvent pas avoir beaucoup d'effet sur les résultats économiques de cette année[41]; (ii) les implications des conditionnalités attachées à un programme du FMI peuvent prendre un certain temps avant de se propager dans l'économie. En particulier, il est possible que les prêts et surtout les conditions attachées soient associées à un effet courbe en J, où l'effet immédiat est négatif, du fait des mesures d'austérité requises, mais l'économie rebondit avec succès lorsque les réformes commencent à avoir un impact positif sur la croissance. D'où la nécessité de tenir compte du délai afin de limiter le biais d'instantanéité dans l'examen des implications macroéconomiques de l'assistance du FMI. Pour tenir compte du biais d'instantanéité (effets instantanés *versus* effets décalés), nous permettons aux programmes du FMI d'avoir un effet retardé sur la croissance.

Notre analyse est fondamentalement basée sur l'estimation d'un modèle de croissance de Solow augmenté (cf. Mankiw et al. 1992). À l'origine, le modèle estimé par Mankiw et al. (1992, p. 411) est:

$$\ln\left(\frac{Y}{L}\right) = a + \frac{\alpha}{1-\alpha}\ln(s_k) - \frac{\alpha}{1-\alpha}\ln(n+g+\delta) + \epsilon, \tag{1}$$

où $Y$ dénote l'output, $L$ le facteur travail, $s_k$ la fraction du revenu investi dans le capital physique, $n$ le taux de croissance du facteur travail, $g$ le taux de croissance du niveau de la technologie (et donc, $n+g$ dénote le taux de croissance du nombre d'unités efficientes du facteur travail), $\delta$ le taux de dépréciation du facteur capital et $\epsilon$ le terme de l'erreur qui est supposé être indépendant et identiquement distribué. En comparant les équations (7), (10) et (12) telles que reprises dans Mankiw et al. (1992), il suit que le terme de l'erreur dans l'équation (1) ci-dessus est tel que:

$$\epsilon = \frac{\beta}{1-\alpha}\ln\left(\frac{s_k^{\alpha} \times s_h^{1-\alpha}}{n+g+\delta}\right)^{1/(1-\alpha-\beta)} + \eta, \tag{2}$$

où $s_h$ dénote la fraction du revenu investi dans le capital humain. Pour estimer notre modèle, nous utilisons les données de la RDC. Les données proviennent de la Banque mondiale (cf. Indicateurs du développement dans le monde, WDI DataBank 2019) et couvrent la période 1960-2018. Nous nous référons à Mankiw et al. (1992, pp. 412-413) pour établir la correspondance entre les variables du modèle et leur équivalent sur données (cf. Tableau 4). La figure 3 propose un aperçu graphique de toutes les variables, y compris les instruments. La variable dépendante, PIB réel divisé par la population en âge de travailler, entre dans le modèle en différence logarithmique. En plus du ratio investissement/PIB et de la croissance de la population en âge de travailler comme variables explicatives[42], nous incluons l'indicateur du programme de FMI, la principale variable explicative.

---

[41] C'est le cas notamment du programme de référence entre le FMI et la RDC (SMP) et de la Facilité rapide de crédit (FRC) validés en décembre 2019.
[42] En référence à Mankiw et al. (1992), voir aussi Barro et Lee (2005, p. 1261), nous utilisons le ratio investissement/PIB plutôt que le ratio stock de capital physique/PIB dans le modèle de Solow augmenté.

Figure 3: Aperçu graphique des variables d'intérêt

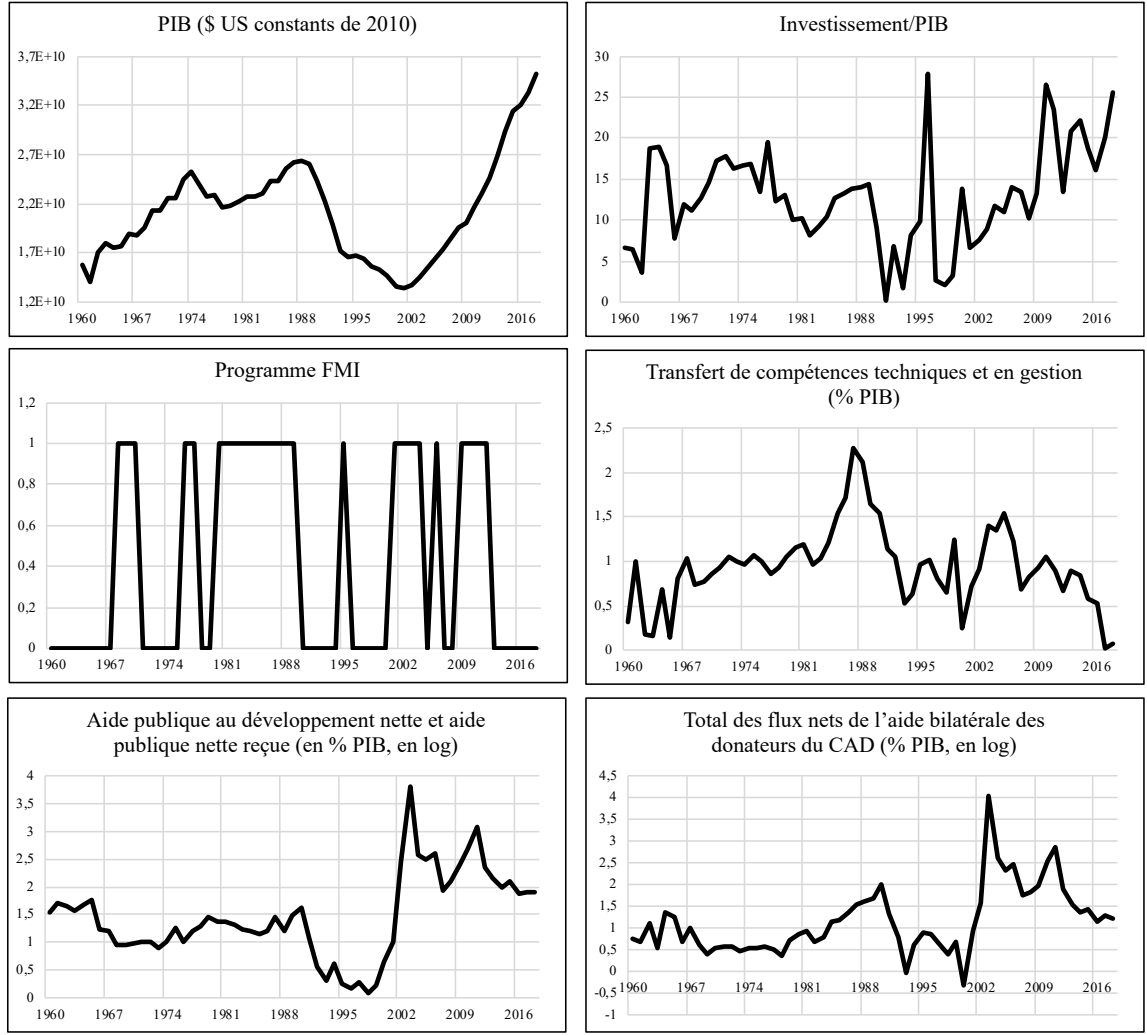

La prise en compte explicite de la variable Programme économique du FMI implique que :

$$\ln\left(\frac{Y}{L}\right) = a + \frac{\alpha}{1-\alpha}\ln(s_k) - \frac{\alpha}{1-\alpha}\ln(n + g + \delta) + \gamma\Phi_{FMI} + \epsilon^*, \qquad (3)$$

où $\gamma$ est un réel et $\epsilon^* = \epsilon - \gamma\Phi_{FMI}$. La variable Programme économique du FMI ($\Phi_{FMI}$) est construite de manière à prendre la valeur de 1 dans les années au cours desquelles le pays est en programme formel avec le FMI et 0 dans le cas contraire. Plus précisément, la variable dummy prend la valeur 1 si le programme était en vigueur pendant au moins 6 mois (cf. Tableau 2 dans la section précédente). Par ailleurs, nous modifierons, une de fois

---

Le recours au ratio investissement/PIB au détriment du ratio stock de capital physique/PIB permet d'atténuer la sensibilité des résultats par rapport aux transformations des données, et donc de préserver une plus grande plausibilité dans l'analyse et l'interprétation des résultats. Voir Kabuya Kalala et al. (2019) pour une discussion détaillée sur la construction de la variable stock de capital.

de plus, l'équation (3) pour évaluer également l'effet sur la croissance économique de l'interaction entre, d'une part, l'aide publique au développement et les programmes économiques du FMI et, d'autre part, les flux nets de l'aide bilatérale des donateurs du Comité d'aide au développement (CAD) et les programmes économiques du FMI. L'aide publique au développement (APD) nette désigne les décaissements de prêts consentis à des taux concessionnels et les subventions des agences membres du CAD, des institutions multilatérales et des pays non-membres du CAD. En revanche, les flux nets d'aide bilatérale des donateurs du CAD dénotent les décaissements nets de l'aide publique au développement (APD) ou de l'aide officielle des membres du CAD.

Tableau 5: Corrélations contemporaines et décalées avec le PIB per capita (en %)

|  | Retard: 0 | Retard: 1 | Retard: 2 | Retard: 3 |
|---|---|---|---|---|
| Programme FMI | -2,62 | -0,85 | -0,94 | -0,08 |
| Investissement/PIB | 9,79 | 15,71 | 21,35 | 24,34 |
| APD (en % PIB) | -31,92 | -25,70 | -19,86 | -15,21 |
| Aide bilatérale (% PIB) | -47,35 | -44,18 | -42,09 | -40,95 |

Le tableau 5 présente les corrélations contemporaines (cf. Retard: 0) et décalées entre le PIB per capita en RDC et les variables d'intérêt de notre analyse, à noter le programme d'assistance du FMI, le ratio Investissement/PIB et les aides. Il ressort une corrélation négative entre le PIB per capita et le programme d'assistance du FMI, quel que soit le décalage considéré[43]. Ces corrélations sont faibles mais statistiquement significatives au seuil de 5 pour cent. La même nature de lien (corrélation négative) est également observée mais avec un degré d'intensité plus élevé entre, d'une part, le PIB per capita et l'aide publique au développement et, d'autre part, le PIB per capita et les flux nets d'aide bilatérale. Autrement dit, ces résultats suggèrent que les interactions entre, d'une part, les programmes du FMI et les aides au développement respectivement et, d'autre part, les performances économiques en RDC sont significativement négatives (*valeur p < 0.05*)[44].

---

[43] Notons que l'association entre le PIB per capita et le programme du FMI mesure le coefficient d'interception différentielle obtenu en appliquant une régression linéaire simple comme suit PIB = $\alpha$ + $\beta$*FMI (cf. Wooldridge 2009, pp. 225-233). La principale interprétation que l'on puisse faire dans ce cas est que lorsque le coefficient $\beta$ est positif ($\beta$>0), cela suggère que le PIB per capita a tendance à être plus élevé en présence du programme du FMI qu'en son absence. En revanche, lorsque le coefficient $\beta$ est négatif ($\beta$<0), cela suggère que le PIB per capita a tendance à être plus faible en présence du programme du FMI qu'en son absence.

[44] Comme le notent Burnside et Dollar (2000), puis Easterly (2003), l'aide étrangère a un impact positif sur la croissance économique dans les pays en développement lorsque la gestion des finances publiques est saine, et lorsque les politiques monétaires et commerciales sont appropriées. Autrement dit, en

Tableau 6: Effets contemporains et décalés des programmes du FMI sur le PIB per capita
(semi-élasticité, 1961-2018)

| | 1961-2018 | 2001-2018 | 1961-2018 | 2001-2018 | 1961-2018 | 2001-2018 |
|---|---|---|---|---|---|---|
| FMI | -0.014 | -0.002 | -0.025 | -0.011 | -0.040 | -0.021 |
| FMI (-1) | | | -0.059 | -0.004 | | |
| FMI (-2) | | | -0.017 | -0.010 | | |
| FMI (-3) | | | -0.005 | -0.004 | | |
| INVEST/PIB | 0.042 | 0.007 | 0.122 | 0.044 | 0.041 | 0.028 |
| Démographie | -0.034 | -0.041 | -0.152 | -0.065 | -0.002 | -0.051 |
| FMI*APD | | | | | -0.047 | -0.008 |
| FMI*AB | | | | | -0.007 | -0.004 |
| Constante | 0.017 | -0.147 | 0.026 | -0.023 | -0.023 | -0.199 |
| SCT | oui | oui | oui | oui | oui | oui |
| SHCT | oui | oui | oui | oui | oui | oui |
| PROB(J-STAT) | 0.471 | 0.752 | 0.663 | 0.649 | 0.479 | 0.550 |

Note: Les estimations pour les variables FMI, ratio investissement/PIB, Démographie dans les différentes spécifications sont toutes statistiquement significatives, au pire, au seuil de 5%. La variable dépendante est le PIB per capita, exprimé en différence logarithmique. Pour l'estimateur de la méthode des moments généralisés soit identifié, il doit y avoir au moins autant d'instruments qu'il y a de paramètres dans le modèle. Ainsi, nous avons introduit des retards dans nos instruments pour respecter cette contrainte. L'estimateur robuste de White a été utilisé. Toutefois, nos résultats demeurent qualitativement non sensibles au choix de la matrice de pondération. Enfin, pour les différents exercices, la probabilité critique associée à la valeur de la fonction objectif optimisée (i.e. statistique J) s'est avéré statistiquement non significative.

---

présence de politiques médiocres, les effets de l'aide étrangère peuvent s'avérer faiblement positifs, non significatifs et même négatifs (cf. Mahembe et Odhiambo 2020).

À présent, nous procédons à l'évaluation de l'impact des programmes économiques du FMI et des aides au développement sur la croissance économique. Le tableau 6 présente les résultats de l'estimation du modèle par la méthode des moments généralisés. Comme nous l'avons mentionné précédemment, nous recourrons à la méthode des moments généralisés pour prévenir les éventuels biais que peuvent induire le caractère sélectif des programmes d'assistance du FMI. Le PIB par habitant est mesuré en différence logarithmique. La première colonne présente les résultats avec la variable FMI entrant de manière contemporaine dans la régression. Cette régression confirme le résultat général suggérant que l'effet du programme d'assistance du FMI sur la croissance économique est négatif et statistiquement significatif au seuil de 5%[45]. Les deux autres variables explicatives sont également significatives au seuil de 5%, l'investissement ayant un effet positif alors que celui de la croissance démographique est négatif.

Pour examiner comment le taux de croissance se comporte au fil du temps par rapport au programme d'assistance du FMI, nous retardons la variable FMI de 1 à 3 ans (colonnes 2, 3 et 4). Cela n'entraîne pas de changement majeur dans les résultats. En effet, l'effet de la participation à un programme du FMI sur la croissance économique demeure négatif et statistiquement significatif. Par ailleurs, les effets des deux autres variables explicatives restent inchangés. Enfin, les colonnes 5 et 6 intègrent dans le modèle les variables d'interaction entre le programme du FMI et l'aide au développement. Cette régression confirme, une fois de plus, le résultat général suggérant que l'effet du programme d'assistance du FMI sur la croissance économique est négatif et statistiquement significatif. Dans les différents exercices, la probabilité critique associée à la valeur de la fonction objectif optimisée est demeurée statistiquement non significative. Ainsi sommes-nous assez confiants que nos instruments sont valables.

## 4. Discussions

Les différentes estimations économétriques suggèrent que les programmes d'assistance technique ou financière mis en œuvre par le FMI en RDC, entre 1960 et 2018, ont eu une incidence négative sur le profil de croissance économique. Notre analyse s'inscrit donc dans la continuité des travaux antérieurs qui ont trouvé des preuves d'un effet négatif ou non significatif des programmes du FMI sur la croissance économique des pays bénéficiaires[46]. De ce fait, devrions-nous conclure que l'assistance du FMI une malédiction pour les bénéficiaires?

---

[45] Toute chose étant également par ailleurs, il ressort que l'effet des programmes du FMI sur le PIB per capita est négatif. Voir Wooldridge (2009, pp. 232) pour l'interprétation d'un coefficient associé à une variable binaire lorsque la variable dépendante est exprimée en logarithme.

[46] Ces résultats renforcent en quelque sorte la thèse soutenue par Stiglitz (2002), dans son livre *Globalization and its Discontents*, stipulant que les politiques préconisées par le FMI entraînant des conséquences sociales dévastatrices et un accroissement de la pauvreté. En effet, en référence à son expérience d'économiste en chef à la Banque mondiale entre 1997 et 2001, Stiglitz soutient que les actions du FMI lors de la crise asiatique, de la transition des ex-pays communistes vers l'économie de marché ou des problèmes de liquidité de pays en développement ont eu une influence néfaste sur les performances économiques des pays concernés en aggravant les crises économiques ou en déclenchant des crises sociales.

Figure 4: Exécution du programme du FMI et pris en en compte du binôme leadership-gouvernance

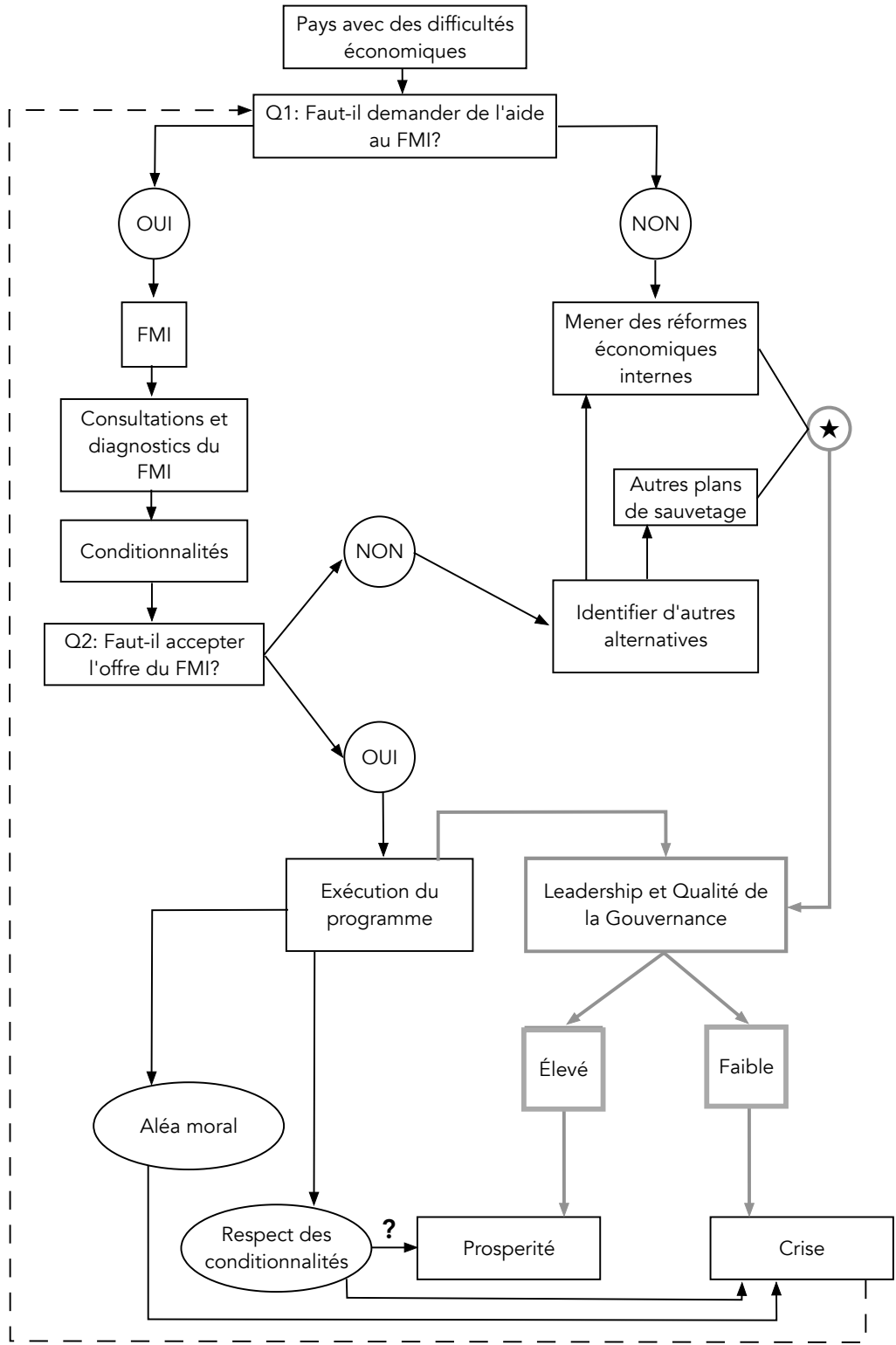

Nous disons: non. En effet, la littérature économique est d'avis qu'une assistance du FMI est susceptible d'affecter négativement le profil de croissance économique du pays bénéficiaire, pour au moins deux raisons : du fait, d'une part, de l'aléa moral et, d'autre part, du syndrome hollandais. Dans ce papier, nous soutenons que, dans la plupart des cas, l'échec des programmes du FMI est expliqué par un déficit du leadership et de la gouvernance[47]. Autrement dit, la nature de l'effet des programmes du FMI sur la croissance économique des pays bénéficiaires est conditionnelle au binôme leadership-gouvernance. L'argumentaire que nous développons, pour soutenir notre thèse, se résume comme suit[48].

Étape 0. Supposons qu'un pays quelconque est en proie à des difficultés économiques.

Étape 1. Éventuellement, l'une des questions que se poseront les autorités politiques est:
– Que faut-il faire pour endiguer, au plus vite, la crise qui sévit?
– Faut-il solliciter l'assistance du FMI?

Étape 2. La réponse à cette dernière question peut être « oui » ou « non ».

Étape 3b. Au cas où la réponse serait « non », il faudra :
– soit envisager de mener des réformes avec les moyens internes.
– soit identifier des plans de sauvetages alternatifs, i.e. plans de sauvetage autres que les programmes économiques du FMI.

Étape 3a. En revanche, au cas où la réponse serait « oui », il faudra donc entamer les négociations avec le FMI.

Étape 4a. À l'issue des négociations avec le FMI, il faudra répondre à une autre question:
– Le pays est-il d'accord avec l'offre du FMI?
– Sommes-nous prêts à respecter les conditionnalités imposées par le FMI?

*Comme précédemment, la réponse à cette question peut être oui ou non.*

Étape 5b. Au cas où la réponse serait « non », il faudra:
– soit envisager de mener des réformes avec les moyens et ressources (humaines, techniques et financières) internes.
– soit identifier des plans de sauvetages alternatifs, i.e. autres que les programmes économiques du FMI.
 *Autrement dit, retour à l'étape 3b.*

---

[47] Pour raison de commodité analytique, nous définissons le leadership comme la volonté et la capacité pour un décideur politique de résister ou de négocier stratégiquement avec les groupes de pression avec comme objectif majeur: prioriser l'atteinte des objectifs favorables au bien-être collectif. La gouvernance est vue comme étant la volonté et la capacité pour un décideur politique de conduire avec succès des réformes structurelles favorables au bien-être collectif. Cf. Annexe F pour une brève discussion sur la définition explicite du leadership éclairé et de la gouvernance de qualité.
[48] Cf. Figure 4 pour un aperçu graphique.

Étape 5a. En revanche, au cas où la réponse serait « oui », le programme d'assistance technique ou financière proposé par le FMI peut donc être mis en œuvre

Étape 6. La mise en œuvre du programme du FMI ne garantit pas forcement le regain de la stabilité macroéconomique et le retour d'une croissance économique positive stable et vigoureuse. En réalité, la mise en œuvre du programme économique peut avoir une incidence positive ou négative:

– D'après la littérature économique, deux facteurs peuvent expliquer l'incidence négative du programme du FMI sur le profil de croissance économique du pays bénéficiaire: (i) l'aléa moral; (ii) le syndrome hollandais.

– Par ailleurs, on s'attend à ce que le résultat escompté soit positif au cas où tous les critères quantitatifs imposés pas le FMI étaient respectés.

– Cependant, en plus des critères quantitatifs, il y a également des critères qualitatifs dont l'interprétation peuvent changer selon les contextes politiques ou selon les intérêts de groupes de pression (lobbying); cf. Marysse (2010); Presbitero et Zazzaro (2012); Dreher et al. (2015).

– En outre, il convient de rappeler qu'en général, comme d'ailleurs cela est documenté dans la littérature et comme le suggèrent nos résultats économétriques, l'incidence des programmes d'assistance du FMI sur le profil de croissance des pays bénéficiaires reste mitigée: elle est généralement négative ou neutre (non significative) et rarement positive. D'où, la pertinence de l'étape 7.

Étape 7. En vue d'atténuer le caractère mitigé de l'effet des programmes du FMI sur le profil de croissance, nous proposons que le FMI insiste davantage sur l'offre du leadership et de la gouvernance, et donc de s'y appuyer lors de l'implémentation de ses plans de sauvetage. Cela renforcerait, entre autres, l'appariement entre la quantité des fonds injectés dans le cadre des plans de sauvetage et la qualité des gestionnaires locaux[49].

De ce qui précède, il convient de noter, pour raison de commodité analytique (cf. Annexe F pour une discussion), que le leadership est défini ici comme la volonté et la capacité pour un décideur politique de résister ou de négocier stratégiquement avec les groupes de pression avec comme objectif majeur: prioriser l'atteinte des objectifs favorables au bien-être collectif. Par ailleurs, la gouvernance doit être vue comme la volonté et la capacité pour un décideur politique de conduire avec succès des réformes structurelles ou transformationnelles favorables au bien-être collectif. Dès lors estimons-nous, en référence notamment aux étapes 6 et 7, qu'à chaque fois que dans un pays donné les piliers du leadership et de la gouvernance seront fragiles, les performances économiques demeureront médiocres, indépendamment du fait que le pays soit en programme ou non avec le FMI. Pour illustrer le lien entre les performances économiques et le binôme leadership-gouvernance, considérons l'indice de perception de la gouvernance de la politique budgétaire (cf. Figure 5). Les niveaux les plus élevés de la qualité de la gouvernance dans la gestion de la politique budgétaire entre 2000 et 2018 ont été observés respectivement: (i) en 2002 au moment de la reprise du programme avec le FMI (PIR) et

---

[49] Voir aussi Hülsmann (2003) pour une discussion détaillée à ce sujet.

de la mise en œuvre du PEG-1; (ii) en 2010 au moment de la mise en œuvre de la Troïka politique par le ministre des finances –un mécanisme de coordination et de suivi régulier des politiques budgétaire et monétaire; et (iii) en 2012 au moment de la mise en œuvre de la Troïka politique par le premier ministre, –un mécanisme de coordination et de suivi régulier de la politique économique du gouvernement.

Figure 5: Perception de la gouvernance de la politique budgétaire en RDC avec ou sans l'assistance du FMI

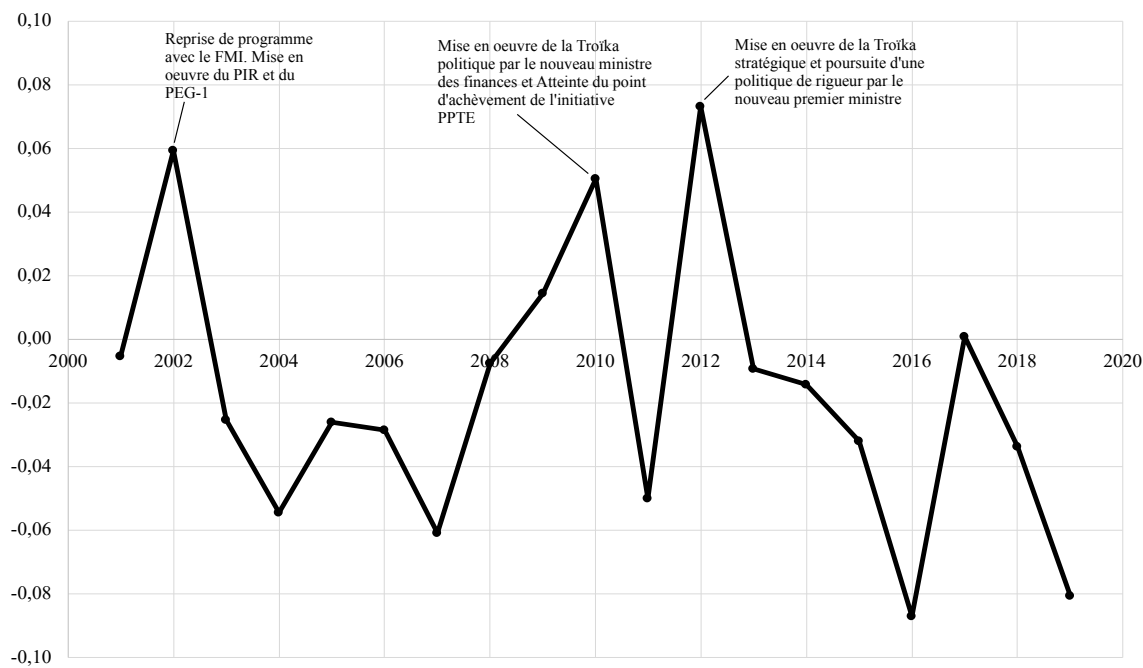

Note: L'indice de perception de la gouvernance de la politique budgétaire mesure le rapport entre le solde des opérations des administrations publiques et les recettes publiques totales. Naturellement, cet indice dénote le degré d'implication et de responsabilité d'un gouvernement dans la maîtrise à la fois des dépenses publiques et du déficit public. Les détails explicatifs et autres indications sur les différentes sources des données sont repris en Annexe A (Voir Tableau B.1 repris en Annexe B).

Remarquons ici qu'à la gouvernance, il faudra un leadership affirmé pour lutter véritablement contre les pesanteurs (ou plus explicitement contre les lobbys ou groupes de pression) qui préfèrent le statu quo plutôt que la mise en œuvre des réformes courageuses et audacieuses[50]. Autrement dit, sans un leadership affirmé (et éclairé), la gouvernance, peu importe sa qualité, sera toujours étouffée par les groupes de pressions internes et externes[51].

Figure 6: Moyenne annuelle des indices EPIN en RDC entre 2005 et 2018

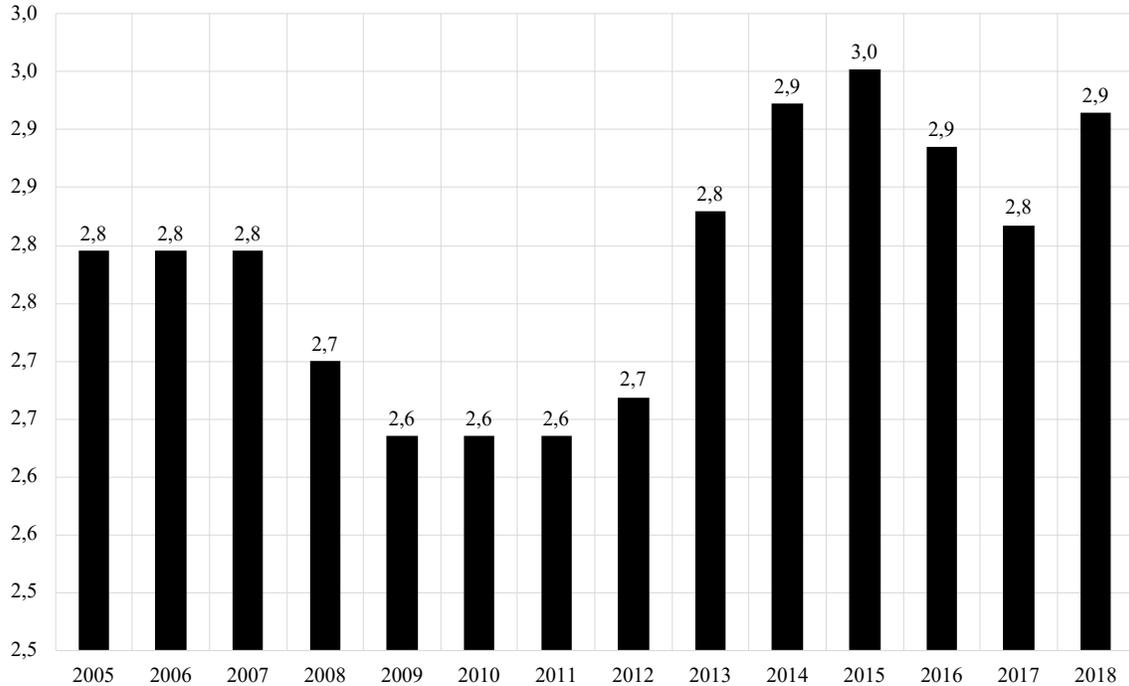

Les indices d'évaluation de la qualité des politiques et des institutions nationales (EPIN) classe les pays en fonction d'un ensemble de critères regroupés en quatre blocs: (i) Gestion économique; (ii) Politiques structurelles; (iii) Politiques d'inclusion sociale et d'équité; et (iv) Gestion et les institutions du secteur public. Les performances d'un pays sont notées sur une échelle allant de 1 (faibles ou médiocres) à 6 (élevées). Source : Banque mondiale (WDI DataBank 2019). Les données utilisées pour élaborer cette figure sont reprises en détails en Annexe G.

Enfin, pour compléter l'analyse ci-dessus, nous suivons Bersch et Botero (2014) et utilisons la base des données de la banque mondiale (cf. Base de données EPIN du Groupe de la Banque mondiale, IDA 2017)[52]. À la lumière de ces données, nous décrivons le proxy

---

[50] Par exemple, entre 2013 et 2014, la RDC avait gagné 11 places dans le classement mondial des pays suivant l'indice de développement humain (cf. Rapport IDH 2015). À ce propos, le PNUD a indiqué que jamais la RDC n'avait obtenu un score semblable, même si elle demeure dans la catégorie des pays à IDH faible. Tout le long de ce papier, nous avons soutenu que la réalisation de telle performance n'était possible que conditionnellement au leadership éclairé et à la gouvernance de qualité et, par ailleurs, indépendamment du fait que le pays soit en programme ou non avec le FMI (cf. Étape 7, Figure 4). Cette thèse corrobore les propos tenus par le représentant résident du PNUD en RDC, Dr Mamadou Diallo RDC, lors du lancement national du rapport sus-évoqué: « Ce score a été obtenu à la suite des réformes courageuses et audacieuses entreprises par le gouvernement congolais [...] ».

[51] Voir aussi Trefon (2010) pour une discussion sur la capacité des fonctionnaires désillusionnés d'entraver les réformes structurelles.

[52] Évaluation de la qualité des politiques et des institutions nationales (CPIA, *Country Policy And Institutional Assessment*). Dans Matata et Tsasa (2019), cette base de données a également été utilisée

ou l'équivalent empirique du binôme leadership-gouvernance comme la capacité, pour un décideur politique, de garantir : (1) la qualité de la gestion budgétaire; (2) un niveau minimal acceptable de bien-être social; (3) l'efficacité des politiques structurelles; (4) la promotion de l'égalité d'accès pour les hommes et les femmes à l'éducation, à la santé, à l'économie et à la protection; (5) la qualité de l'administration publique; (6) la stabilité du cadre macroéconomique; (7) la transparence des activités dans le secteur public; (8) un cadre politique favorisant le commerce de biens; (9) la soutenabilité à court et à moyen terme de la politique fiscale; (10) une structure de gouvernance basée sur le respect des droits de propriétés et sur des règles fiables; (11) la fourniture de services par les secteur public et privé qui ont une incidence sur l'accès aux services d'éducation et de soins de santé et sur leur qualité; (12) une stratégie de gestion de la dette favorisant la réduction des risques budgétaires et la soutenabilité de la dette à long terme; (13) un secteur financier efficient; (14) la protection et l'utilisation durable des ressources naturelles et la gestion de la pollution; (15) l'efficacité de la mobilisation des revenus; (16) un environnement qui favorise la création d'emploi et une augmentation de la productivité; (17) les priorités nationales en matière de réduction de la pauvreté; (18) l'inclusion sociale et l'équité de l'utilisation des ressources publiques. Il ressort de la figure 6 que le pic le plus élevé observé en 2012 dans la figure 5 coïncide avec la période où la moyenne annuelle des indices EPIN en RDC a entamé une phase ascendante, passant de 2,6 en 2011 à 2,7 en 2012, avec un pic de 3.0 en 2015. La période où la moyenne annuelle des indices EPIN en RDC est la plus élevée (équivalent approximatif d'un indice binôme leadership-gouvernance élevé) coïncide avec la période 2012-2016, une période où (i) la RDC n'était pas en programme avec le FMI et (ii) le taux de croissance par habitant affiche son niveau le plus élevé (cf. Figure 2). Ainsi pour dire qu'il est bien possible qu'un pays, comme la RDC, réussisse effectivement à s'imposer une « rigidité de règle » telle que requise par les différents partenaires de développement avec l'assistance du FMI (2005-2011) ou sans l'assistance du FMI (entre 2012-2016). Plus important encore, en comparant les deux périodes en termes de croissance du PIB par habitant, il ressort que la croissance a été plus élevée entre 2012 et 2016 qu'entre 2005 et 2011. À notre avis, le facteur déterminant qui explique ce différentiel de performances n'est pas le fait d'être ou non en programme avec le FMI, c'est plutôt la nature du leadership et la qualité de la gouvernance. La figure 6 semble apporter un support empirique supplémentaire à cette assertion.

## 5. Conclusion

Un nombre relativement important d'études empiriques dans la littérature économique ont suggéré qu'en général les effets des programmes du FMI sur la croissance économique sont négatifs ou statistiquement non significatifs. Cependant, pour diverses raisons, ces résultats peuvent ne pas être généralisés, notamment lorsqu'il s'agit de considérer le cas spécifique ou isolé d'un pays. Par exemple, les études existantes varient considérablement en termes de la couverture des programmes considérés, des caractéristiques des pays inclus dans l'échantillon, de la période retenue ou de la méthodologie employée. Aussi, elles varient en termes d'efficacité avec laquelle elles gèrent les potentiels biais d'endogénéité

---

pour examiner la validité empirique de l'hypothèse de la malédiction des ressources naturelles. La description des indices EPIN est reprise en Annexe G.

ou d'instantanéité qu'impliquent ces programmes. D'où, notamment, l'intérêt de procéder à l'étude au cas par cas.

À ce jour, au mieux de notre connaissance, il n'existe quasiment aucune étude qui examine de manière formelle les implications macroéconomiques des programmes du FMI en République démocratique du Congo (RDC). Le présent papier s'est donc proposé de combler tant soit peu ce creux en analysant, sous une perspective historique et prospective, les corrélations contemporaines et retardées entre les programmes d'assistance technique et financière du FMI et les performances économiques en RDC. Nous avons également exploré l'impact de ces programmes sur la croissance économique en instrumentant le transfert de compétences techniques et en gestion. La méthode des moments généralisés a été utilisée afin de tenir compte du potentiel biais d'endogénéité dû au caractère non aléatoire de la sélection du pays qui bénéficie de l'assistance du FMI. La période considérée, à cet effet, est 1961-2018. Nos résultats suggèrent que l'effet contemporain des programmes d'assistance du FMI sur la croissance économique en RDC est négatif et statistiquement significatif au seuil de 5 pour cent. Par ailleurs, dans l'objectif de prendre en compte le potentiel biais d'instantanéité, nous avons examiné comment le taux de croissance en RDC se comporte au fil du temps par rapport aux effets retardés de différents programmes du FMI. Pour ce faire, nous avons introduit un retard de 1 à 3 ans dans la variable FMI. Une fois de plus, les résultats des différentes régressions suggèrent que les effets retardés de la participation à un programme du FMI sur la croissance économique en RDC sont négatifs et statistiquement significatifs. Ces résultats demeurent robustes même lorsque nous considérons la sous-période 2001-2018. i.e. une période plus récente.

À première vue, ces prédictions peuvent paraître surprenantes, en ce sens que le rôle du FMI est d'apporter, le cas échéant, une assistance technique et financière en faveur des pays membres qui éprouvent des difficultés économiques. Cependant, il sied de rappeler que la littérature économique est d'avis qu'une assistance du FMI est susceptible d'affecter négativement le profil de croissance économique du pays bénéficiaire, pour au moins deux raisons: du fait, d'une part, de l'aléa moral et, d'autre part, du syndrome hollandais. Dans ce papier, nous avons plutôt soutenu, sur base de l'expérience de la RDC, que l'échec des programmes du FMI est fondamentalement causé par le déficit, à la fois, du leadership éclairé et de la gouvernance de qualité.

Par ailleurs, nous avons également montré, en nous basant sur une analyse fine des statistiques, que les approches *before-after* et *with-without* qui se focalisent sur le fait d'être ou non en programme avec le FMI ne suffisent pas à expliquer, à elles seules, de manière cohérente le différentiel observé dans les performances économiques en RDC entre 1960 et 2018. De ce qui précède, en nous focalisant sur le cas de la RDC, nous avons apporté quelques preuves empiriques illustrant qu'un pays en développement peut réussir la stabilisation du cadre macroéconomique et l'ajustement économique avec ou sans l'assistance du FMI, *conditionnellement* à la nature du leadership et à la qualité de la gouvernance. Par exemple, durant les périodes 1991-2000 et 2012-2016, la RDC n'était pas en programme avec le FMI. Le taux de croissance du PIB par habitant entre 1991 et 2000 était de -7.89 pour cent, alors qu'entre 2012 et 2016, ce taux se chiffrait à 3,39 pour cent. De même, durant les périodes 1980-1991 et 2001-2012, la RDC était en programme avec le FMI. Pourtant le taux de croissance du PIB par habitant entre 1980 et 1991 était de -1,66 pour cent, alors qu'entre 2000 et 2011, ce taux se chiffrait à 1,63 pour cent. Les différentes analyses fines des statistiques que nous avons réalisées tendent à attribuer, de manière convergente et concordante, ce différentiel de performances à la nature du leadership et à la qualité de la gouvernance. En outre, cette conclusion corrobore également les évidences empiriques sur la comptabilité des réformes qui révèle qu'entre 2001 et 2016, et plus particulièrement entre 2009 et 2016, plusieurs mécanismes institutionnels exceptionnels furent mis en œuvre, à l'instar notamment de la troïka stratégique, de la troïka politique, de la cotation des ministres du gouvernement central, de la réforme de l'administration publique ou de l'exécution programmatique des travaux publics. L'objectif de ces mécanismes était de procéder, de manière systématique, à une surveillance continue du cadre macroéconomique et à un suivi rigoureux des réformes structurelles. Cependant, force est de constater que, depuis 2016, l'effectivité de ces mécanismes s'est vue quelque peu affaiblie par la succession des gouvernements de cohabitation et de coalition qui tendent à prioriser les arrangements politiques, au détriment de l'efficacité socio-économique. À cet effet, nous soutenons et rappelons, à cet effet, que le développement économique et le progrès social de la RDC se fonderont non pas sur la nécessité de pouvoir garantir, chaque fois, le consensus des classes politiques (lequel s'accompagne souvent avec des gouvernements pléthoriques), mais plutôt sur l'impérative de consentir des sacrifices durant plusieurs années successives en vue de promouvoir des politiques et projets dont la qualité du contenu et l'articulation dans leurs mises en œuvre sont cohérentes avec les objectifs de croissance. Il est clair que cette option dépend donc de manière critique de la qualité du leadership de la classe dirigeante, en ce sens que le leadership politique dénote la capacité, à la fois, (i) d'avoir une vision claire du progrès de son pays, partagée par l'ensemble ou la majeure partie de la population, (ii) de définir des objectifs globaux et intermédiaires précis et les mettre en œuvre, (iii) de résister aux pesanteurs de toute nature pouvant fragiliser ou anéantir la démarche entreprise et (iv) de vaincre les obstacles érigés par les groupes de pression et lobbyistes.

## Bibliographie

**ANNEXES**

**Annexe A. Données, sources et informations complémentaires**

Tableau A.1. Différentiel des performances économiques de la RDC par rapport à celles de l'Afrique subsaharienne

|  | 1980-1990 | 1991-2000 | 2001-2011 | 2012-2016 | 2017-2018 |
|---|---|---|---|---|---|
| Différentiel de croissance | -0,44 | -7,71 | -0,93 | 2,60 | 1,63 |
| *Taux de croissance en RDC* | *-1,66* | *-8,36* | *1,63* | *3,39* | *1,40* |
| *Taux de croissance en Afrique subsaharienne* | *-1,21* | *-0,64* | *2,56* | *0,79* | *-0,23* |
| Différentiel d'inflation | 56,73 | 3891,62 | 18,20 | -1,04 | 32,64 |
| *Taux d'inflation en RDC* | *66,80* | *3901,35* | *25,07* | *2,58* | *36,60* |
| *Taux d'inflation en Afrique subsaharienne* | *10,06* | *9,73* | *6,87* | *3,62* | *3,96* |
| Assistance technique et financière du FMI | oui | non | oui | non | non |

La croissance en RDC nette de l'Afrique subsaharienne mesure la différence entre le taux de croissance du PIB par habitant en RDC et la moyenne du taux de croissance du PIB par habitant au niveau de l'Afrique subsaharienne. Pour une période donnée, une valeur positive de la croissance en RDC nette de l'Afrique subsaharienne signifie que le taux de croissance économique réalisé en RDC est supérieur à la moyenne du taux de croissance enregistrée au niveau de l'Afrique subsaharienne. En revanche, une valeur négative de la croissance en RDC nette de l'Afrique subsaharienne signifie que le taux de croissance économique observé en RDC est inférieur à la moyenne du taux de croissance enregistrée au niveau de l'Afrique subsaharienne. En parallèle, l'inflation en RDC nette de l'Afrique subsaharienne mesure la différence entre le taux d'inflation en RDC et la moyenne du taux d'inflation au niveau de l'Afrique subsaharienne. Pour une période donnée, une valeur positive de l'inflation en RDC nette de l'Afrique subsaharienne signifie que le taux d'inflation réalisé en RDC est supérieur à la moyenne du taux d'inflation enregistrée au niveau de l'Afrique subsaharienne. En revanche, une valeur négative de l'inflation en RDC nette de l'Afrique subsaharienne signifie que le taux d'inflation observé en RDC est inférieur à la moyenne du taux d'inflation enregistrée au niveau de l'Afrique subsaharienne. Les données sur le taux de croissance du PIB par habitant et le taux d'inflation (déflateur du PIB, RDC et Afrique subsaharienne) proviennent de la base des Indicateurs du développement dans le monde de la Banque mondiale (WDI DataBank 2019).

Tableau A.2. Élasticité du PIB par habitant en RDC par rapport au prix des matières premières

| | 1980-1990 | 1991-2000 | 2001-2011 | 2012-2016 | 2017-2018 |
|---|---|---|---|---|---|
| Élasticité du PIB par habitant en RDC par rapport à l'indice mondial des prix des métaux (1992=100) | . . . | 0,24 | 0,12 | -0,21 | 0,37 |
| Élasticité du PIB par habitant en RDC par rapport au prix mondial du cuivre (USD/tonne métrique) | . . . | 0,63 | 0,11 | -0,25 | 0,42 |
| *Écart-type du taux de croissance du PIB par habitant* | . . . | *0,23* | *0,08* | *0,06* | *0,02* |
| *Écart-type de l'indice mondial des prix des métaux* | . . . | *0,12* | *0,60* | *0,25* | *0,05* |
| *Écart-type du prix mondial du cuivre* | . . . | *0,19* | *0,67* | *0,21* | *0,04* |
| *Corrélation entre taux de croissance et prix des métaux* | . . . | *0,12* | *0,97* | *-0,89* | *1,00* |
| *Corrélation entre taux de croissance et prix du cuivre* | . . . | *0,51* | *0,94* | *-0,88* | *1,00* |
| Assistance technique et financière du FMI | oui | non | oui | non | non |

Dans le papier, l'élasticité-croissance du prix mondiale du cuivre dénote l'élasticité du PIB par habitant (évaluée en logarithme) en RDC par rapport au prix mondial du cuivre (également évalué en logarithme). Par construction, l'élasticité-croissance du prix mondiale du cuivre précise dans quelle mesure les variations du prix du cuivre sur le marché mondial des matières premières affectent les performances économiques en termes de croissance économique. Elle est calculée comme suit (écart-type du log du PIB par habitant/écart-type du log du prix mondial du cuivre)×(corrélation entre log du PIB par habitant et log du prix mondial du cuivre). (i) Les données sur le PIB par habitant proviennent de la base des Indicateurs du développement dans le monde de la Banque mondiale (WDI DataBank 2019). (ii) Les données sur le prix mondial du cuivre proviennent de la Réserve fédérale de Saint-Louis. Même commentaire pour L'élasticité du PIB par habitant en RDC par rapport à l'indice mondial des prix des métaux.

Tableau A.3. Solde de la balance générale des paiements (en % PIB)

| | 1980-1990 | 1991-2000 | 2001-2011 | 2012-2016 | 2017-2018 |
|---|---|---|---|---|---|
| Balance des paiements | -0,52 | -6,95 | -2,89 | 0,11 | -1,40 |
| *Balance commerciale* | *-4,18* | *-1,36* | *-4,91* | *-6,44* | *-3,83* |
| *Compte courant* | *-2,94* | *-3,71* | *-3,40* | *-5,31* | *-3,74* |
| *Compte de capital et des opérations financières* | *2,42* | *-4,22* | *1,02* | *5,53* | *3,28* |
| Assistance technique et financière du FMI | oui | non | oui | non | non |

Note: Les données sur les soldes de la balance des paiements, de la balance commerciale, du compte courant, et du compte de capital et des opérations financières proviennent de la Banque centrale (BCC). Les données pour la période 1987-1994 sont tirées du Tableau III.29 du Rapport annuel 2007 de la BCC (page 147). Pour la période 1995-1999, cf. Tableau III.29bis du Rapport annuel 2007 de la BCC (page 148). Pour la période 2003-2004, cf. Tableau III.26 du Rapport annuel 2008 de la BCC (page 150). Pour la période 2005-2007, cf. Tableau III.28 du Rapport annuel 2013 de la BCC (page 167). Pour la période 2008-2017, cf. Tableau III.29 du Rapport annuel 2017 de la BCC (page 121). Les données pour l'exercice 2018 proviennent de la base des Indicateurs du développement dans le monde de la Banque mondiale. En référence au cinquième manuel de la balance des paiements du FMI: *Solde Global de la balance des paiements* = Solde du Compte courant + Solde du Compte de capital et d'opérations financières + Erreurs et omissions. Par ailleurs: (i) *Solde du Compte courant* = Solde Balance commerciale (Compte des Biens et Services) + Solde Compte de Revenus + Compte de Transferts Courants; (ii) *Solde du Compte de capital et d'opérations financières* = Solde Compte de Capital + Solde Compte d'Opérations financière.

**Annexe B. Gouvernance de la politique budgétaire[53]**

À l'effet de calculer l'indice de perception de la gouvernance de la politique budgétaire (cf. Tableau B.1), pour chaque période, nous divisons le solde des opérations des administrations publiques (exprimés en millions de CDF) par les recettes publiques totales (exprimées en millions de CDF). Le solde des opérations des administrations publiques mesure la différence entre les recettes publiques et les dépenses publiques. Naturellement, l'indice de perception de la gouvernance de la politique budgétaire mesure donc le degré d'implication et de responsabilité d'un gouvernement dans la maîtrise à la fois des dépenses publiques et du déficit public. Lorsque cet indice tend vers des valeurs positives élevées, le gouvernement est perçu plus rigoureux. En revanche, lorsque l'indice tend vers des valeurs négatives élevées en valeur absolue, le gouvernement est perçu moins rigoureux. Un gouvernement est justement rigoureux lorsque l'indice prend la valeur zéro.

---

[53] Ce texte est un extrait tiré de Matata et Tsasa (2020).

Tableau B.1. Perception de la gouvernance de la politique budgétaire

| | Solde des opérations des administrations publiques (millions CDF) | Recettes publiques (millions CDF) | Perception de la gouvernance de la politique budgétaire |
|---|---|---|---|
| 2001 | -796.97 | 148 551.29 | -0.01 |
| 2002 | 17 607.90 | 297 200.98 | 0.06 |
| 2003 | -11 219.10 | 445 225.39 | -0.03 |
| 2004 | -29 205.26 | 536 322.00 | -0.05 |
| 2005 | -29 706.27 | 1 139 225.21 | -0.03 |
| 2006 | -34 708.75 | 1 216 224.25 | -0.03 |
| 2007 | -55 269.40 | 906 981.60 | -0.06 |
| 2008 | -16 033.00 | 2 118 576.40 | -0.01 |
| 2009 | 45 277.00 | 3 134 561.60 | 0.01 |
| 2010 | 152 303.00 | 3 023 273.10 | 0.05 |
| 2011 | -196 895.00 | 3 948 621.80 | -0.05 |
| 2012 | 390 639.00 | 5 331 527.00 | 0.07 |
| 2013 | -50 353.90 | 5 409 849.70 | -0.01 |
| 2014 | -85 945.30 | 6 022 222.90 | -0.01 |
| 2015 | -183 408.40 | 5 725 725.90 | -0.03 |
| 2016 | -431 105.00 | 4 962 421.10 | -0.09 |
| 2017 | 6 257.00 | 6 602 083.30 | 0.00 |
| 2018 | -238 556.89 | 7 086 585.49 | -0.03 |
| 2019 | -553 581.58 | 6 872 806.86 | -0.08 |

Source: (i) Les données sur le solde des opérations des administrations publiques et les recettes publiques totales proviennent de la Banque centrale du Congo (BCC). (ii) Les données pour la période 2001-2007 proviennent du Tableau II.8 du Rapport annuel 2007 de la BCC (page 71); voir aussi Tableau II.9 du Rapport annuel 2008 (page 82) ou encore Tableau II.10 du Rapport annuel 2009 (page 94). (iii) Les données pour la période 2008-2017 proviennent du Tableau II.7 du Rapport annuel 2017 de la BCC (page 61); voir aussi Tableau II.8 du Rapport annuel 2015 (page 84) ou encore Tableau II.8 du Rapport annuel 2016 (page 90). (iii) Les données pour la période 2018-2019 proviennent du Tableau III.2 du no. 51 du Condensé hebdomadaire d'informations statistiques produit par la Direction de la Recherche et des Statistiques de la BCC (Décembre 2019, page 25); voir aussi Tableau III.2 du no. 36 du Condensé hebdomadaire d'informations statistiques produit par la Direction de la Recherche et des Statistiques de la BCC (Septembre 2019, page 25). Les statistiques pour la période 2001-2018 sont des données conciliées de la BCC et du Ministère des Finances. Les données pour l'exercice 2019 concernent la période du 1er janvier 2019 au 20 décembre 2019.

**Annexe C. Quelques informations supplémentaires sur le FMI et ses mécanismes d'intervention**

**C.1. Les instruments utilisés par le FMI**

Le Fonds monétaire international (FMI) emploie fondamentalement deux types de mécanismes de financement ou instruments des prêts, à noter: les instruments de prêt non concessionnel et les instruments de prêt concessionnel (FMI 2015, p.2). Les instruments de prêt non concessionnel comprennent principalement: (i) les accords de confirmation (FMI 2016j), pour surmonter des difficultés temporaires de balance des paiements; (ii) la ligne de crédit modulable (LCM, FMI 2016k), pour les pays ayant des fondamentaux et des antécédents économiques solides; (iii) la ligne de précaution et de liquidité (LPL, FMI 2016l), pour les besoins à moyen terme; (iv) le mécanisme élargi de crédit (MEDC, FMI 2018e), pour remédier à des difficultés de moyen à long terme de balance des paiements dues à de vastes distorsions qui nécessitent des réformes économiques fondamentales; (v) l'instrument de financement rapide (IFR, FMI 2016m), pour l'aide d'urgence des pays confrontés à un besoin urgent en matière de balance des paiements. Tous les instruments de prêt non concessionnel sont assortis d'un taux d'intérêt qui tient compte des fluctuations à court terme des taux d'intérêt sur les principaux marchés monétaires internationaux. Le montant maximum qu'un pays peut emprunter au FMI varie en fonction de la quote-part. Les prêts concessionnels ou facilités de financement concessionnel sont octroyés aux pays à faible revenu, en vue de les aider à atteindre une position macroéconomique durable et compatible avec une croissance et une réduction de la pauvreté vigoureuses. Les conditions de financement offrent une plus grande concessionnalité, avec des taux d'intérêt nuls ou proches de zéro. Les instruments de prêt concessionnel comprennent principalement: (i) la facilité élargie de crédit (FEC, FMI 2016n), pour les pays à faible revenu affrontant des problèmes prolongés de balance des paiements; (ii) la facilité de crédit de confirmation (FCC, FMI 2016o), pour les pays à faible revenu ayant des besoins de balance des paiements potentiels ou à court terme; (iii) la facilité de crédit rapide (FCR, FMI 2016p), pour les pays à faible revenu qui font face à un besoin urgent de balance des paiements; (iv) la composante à accès élevé de la facilité de protection contre les chocs exogènes (FCE, FMI 2017c), destinés à soutenir les pays qui font face à des besoins de balance des paiements causés par des chocs soudains et exogènes; (v) le fonds fiduciaire d'assistance et de riposte aux catastrophes (ARC, FMI 2016ac), en vue d'accorder aux pays les plus pauvres et les plus vulnérables frappés par une catastrophe naturelle ou de santé publique aux conséquences désastreuses des dons leur permettant d'alléger leur dette.

**C.2. Les droits de tirage spéciaux**

Les droits de tirage spéciaux (DTS; en anglais: *Special Drawing Rights*, SDR) est un actif de réserve international, créé en 1969 par le FMI, pour compléter les réserves officielles existantes des pays membres. Les DTS peuvent être échangés contre des devises librement utilisables. À compter du 1er octobre 2016, la valeur du DTS repose sur un panier de cinq grandes devises, à noter: le dollar américain (USD), l'euro, le yuan ou renminbi chinois (RMB), le yen japonais et la livre sterling. Pour plus de détails, voir FMI (2016ab, c).

**C.3. Le programme de référence supervisé par le personnel du FMI**

Le programme de référence (en anglais: *Staff-Monitored Program*, SMP) est un accord informel entre les autorités d'un pays et le personnel du FMI, par lequel ce dernier accepte de suivre la mise en œuvre du programme économique des autorités, incluant notamment les réformes qui visent à augmenter les recettes publiques, à lutter contre la corruption et à améliorer la gouvernance. Le SMP n'entraîne ni assistance financière ni approbation par le Conseil d'administration du FMI.

**Annexe D. Comment fonctionne le FMI[54]**

Le Fonds monétaire international (FMI) et la Banque mondiale, connus sous le nom d'institutions de Bretton Woods, ont été créés en 1944. Le FMI a pour objectif de promouvoir la coopération monétaire internationale, la stabilité des taux de change et l'expansion du commerce international en agissant comme prêteur de dernier recours lorsqu'un pays membre fait face à une crise économique sévère.

En principe, le FMI a une structure semblable à une coopérative financière. Les contributions d'un pays membre au FMI (quotas) sont basées sur son poids dans l'économie mondiale. Ce poids détermine également son pouvoir de vote et sa capacité d'emprunt (tirages). Les quotas équivalent à un échange d'actifs avec peu de coûts directs pour les contribuables. Les cinq principaux contributeurs au FMI sont les États-Unis (16,75%), le Japon (6,23%), l'Allemagne (5,81%), la France (4,29%) et le Royaume-Uni (4,29%)[55].

En s'adressant au FMI, un pays membre confronté à une crise financière a accès aux ressources et aux conseils du fonds. Au fur et à mesure que les tirages d'un pays s'élargissent par rapport à ses quotas, il doit respecter des normes ou des conditionnalités plus strictes, ce qui signifie généralement des changements importants dans les politiques économiques pour garantir que les déficits intérieurs et extérieurs du pays soient considérablement réduits, voire éliminés. Le non-respect de ces conditions entraîne la suspension, la renégociation, voire l'annulation du programme.

Bien que la taille totale des quotas du FMI soit passée d'environ 9 milliards de dollars lors de sa création à près de 677 milliards de dollars américains en 2016. En septembre 2017, 181 des 189 membres avaient effectué leurs paiements de quotas, représentant plus de 99% de l'augmentation totale des quotas, soit environ 675 milliards de dollars.

*In fine*, pour aider à protéger les pays membres qui poursuivent des politiques saines contre les chocs extérieurs et à rehausser la confiance des marchés en temps de risque élevé, le FMI a récemment mis en place la Ligne de crédit modulable (LCM) et la Ligne de précaution et de liquidité (LPL), qui ont essentiellement pour objet de prévenir les crises, mais peuvent également être utilisées pour les résoudre. D'autres instruments nouveaux, comme l'Instrument de financement rapide (IFR) et la Facilité de crédit rapide (FCR) ont également été créés dans le but de fournir une assistance rapide aux pays éprouvant des besoins urgents en matière de balance des paiements, notamment à la suite de chocs engendrés par les prix des produits de base, de catastrophes naturelles et de facteurs de fragilité intérieurs.

---

[54] Texte adapté de De Vries (1987), James (1996), Fischer (1997), Camdessus (1998), Kapur (1998), Mussa et Savastano (1999), Li et al. (2015), et FMI (2016k, 2020).
[55] À noter, ces cinq grandes puissances économiques sont aussi des pays très endettés. Dette publique sur PIB : Japon (234,7%); France (98,2%); Royaume-Uni (90,6%); États-Unis (73,6%); Allemagne (71,7%).

**Annexe E. Le FMI et la croissance économique dans la littérature économique**

La question de savoir si le FMI influe effectivement sur la croissance économique a fait l'objet d'un grand nombre d'études[56]. De manière générale, trois méthodes d'évaluation ont été utilisées dans la littérature.

Tableau E.1: Le FMI et la croissance économique

| Papier/Méthodologie | Période | Nombre de programmes | Nombre de pays | Effet sur la croissance |
|---|---|---|---|---|
| BEFORE-AFTER (Avant/Après) | | | | |
| Reichman et Stillson (1978) | 1963-72 | 79 | ... | Positif |
| Connors (1979) | 1973-77 | 31 | 23 | Aucun |
| Nsouli et Zulu (1985) | 1980-81 | 35 | 22 | Aucun |
| Killick (1986) | 1974-79 | 38 | 24 | Aucun |
| Pastor (1987) | 1965-81 | ... | 18 | Aucun |
| Killik, Malik et Manuel (1992) | 1979-85 | ... | 16 | Positif |
| Rozwadowski et al. (1993) | 1983-93 | 55 | 19 | Positif |
| Evrensel (2002) | 1971-97 | ... | 109 | Aucun |
| Hardoy (2003) | 1970-90 | 460 | 69 | Aucun |
| | | | | |
| WITH-WITHOUT (Avec/Sans) | | | | |
| Donovan (1981) | 1970-76 | 12 | 12 | Positif |
| Donovan (1982) | 1971-80 | 78 | 44 | Négatif |
| Loxley (1984) | 1971-82 | 38 | 38 | Aucun |
| Gylfason (1987) | 1977-79 | 32 | 14 | Aucun |
| Faini et al. (1991) | 1978-86 | ... | 93 | Aucun |
| Hardoy (2003) | 1970-90 | 460 | 69 | Aucun |
| Hutchison (2004) | 1975-97 | 455 | 25 | Aucun |
| Atoyan et Conway (2006) | 1993-2002 | 181 | 95 | Aucun |
| | | | | |
| RÉGRESSION | | | | |
| Goldstein et Montiel (1986) | 1974-81 | 68 | 58 | Aucun |
| Khan (1990) | 1973-88 | 259 | 69 | Négatif |
| Doroodian (1993) | 1977-83 | 27 | 43 | Aucun |
| Conway (1994) | 1976-86 | 217 | 73 | Positif |
| Bagci et Perraudin (1997) | 1973-92 | ... | 68 | Positif |
| Bordo et Schwarz (2000) | 1973-98 | ... | 24 | Négatif |
| Dicks-Mireaux et al. (2000) | 1986-91 | 88 | 74 | Positif |
| Przeworski et Vreeland (2000) | 1970-90 | 465 | 135 | Négatif |
| Hutchison (2003) | 1975-97 | 461 | 67 | Négatif |
| Hutchison et Noy (2003) | 1975-97 | 764 | 67 | Négatif |
| Nsouli, Mourmouras, Atoian (2004) | 1992-2000 | 124 | 92 | Aucun |
| Butkiewicz andYanikkaya (2005) | 1970-99 | 407 | ... | Négatif |
| Barro and Lee (2005) | 1975-99 | 725 | 81 | Négatif |
| Easterly (2005) | 1980-99 | 107 | 107 | Aucun |
| Atoyan and Conway (2006) | 1993-2002 | 181 | 95 | Aucun |

Source : Dreher (2004, p. 31), Dreher (2006, p. 773).

---

[56] C'est à partir des années 1970, que le FMI a commencé à accorder une importance croissante à la croissance économique comme objectif politique (cf. Hardoy 2003, Dreher 2006).

Premièrement, l'analyse *before-after* (avant/après) compare la croissance économique avant l'approbation du programme du FMI avec sa valeur après la période de programmation. Des différences sont ensuite attribuées au programme. Évidemment, cette méthode a ses inconvénients. La participation aux programmes du FMI n'est pas exogène mais est généralement la conséquence d'une crise. En attribuant tous les changements de croissance au cours de la période au programme de FMI fait en sorte que ses effets négatifs sont surestimés. Une deuxième approche utilisée pour évaluer l'impact du FMI sur la croissance a consisté à comparer les taux de croissance dans les pays de programme avec le développement de la croissance dans un groupe témoin (*with-without*). Les chocs exogènes frappant non seulement les pays qui sont programme mais aussi les pays du groupe témoin ne fausseraient pas les résultats. Le problème, bien sûr, est de trouver un groupe témoin adéquat. Idéalement, pour chaque pays qui sont en programme, il devrait y avoir un pays de contrôle exactement dans la même position initiale. Les programmes ne sont cependant pas distribués aléatoirement entre les pays membres, mais sont choisis dans des pays ayant des caractéristiques spécifiques. Comme l'a montré Santaella (1996), la situation initiale des pays qui sont en programme diffère considérablement des pays témoins. Même si le groupe témoin était choisi en fonction d'indicateurs économiques, la différence la plus importante ne pouvait pas être prise en compte, la décision de négocier un programme du FMI en premier lieu. La troisième méthode est l'analyse de régression. Elle a été utilisée par les études les plus récentes. Lorsque l'endogénéité des variables liées au FMI est soigneusement prise en compte, cette méthode semble être la plus prometteuse. Plusieurs études récentes, incluant notamment Przeworski et Vreeland (2000), Barro et Lee (2005) et Drehel (2006) parmi tant d'autres, tiennent compte de l'endogénéité.

Comme nous pouvons l'apercevoir dans le tableau E.2[57], près de 80 pour cent d'études existantes trouvent que les effets des programmes du FMI sur la croissance économique sont soit négatifs, soit statistiquement non significatifs[58]. En revanche, environ 20 pour cent des travaux empiriques trouvent une incidence positive des effets des programmes du FMI sur la croissance économique. Ces éléments contradictoires découlent en partie des différences dans la couverture des pays, les périodes d'échantillonnage et la méthodologie employée.

Tableau E.2: Quels sont les effets des programmes du FMI sur le profil de croissance économique des pays bénéficiaires? Que nous renseigne la littérature?

|  | Effet positif | Effet négatif ou statistiquement non significatif | Total |
|---|---|---|---|
| Nombre de papier | 7 | 25 | 32 |
| En pourcentage (%) | 21,9 | 78,1 | 100 |

In fine, comme nous le montrons dans le texte (cf. section 2), une mise à jour des statistiques reprises dans le tableau E.2 préserve la même tendance, i.e. la plupart des études empiriques, même les plus récentes, tendent à trouver que les implications macroéconomiques des programmes du FMI sont généralement négatives ou statistiquement non significatives.

---

[57] Tableau élaboré sur base d'informations contenues dans le tableau E.1.
[58] Au regard de nouvelles évidences établies par Abadie (2020) et de la prépondérance des résultats non significatifs statistiquement, une analyse plus attentive de ces derniers, sous forme notamment d'une revue de littérature, est une avenue intéressante pour les recherches futures.

**Annexe F. Leadership éclairé et gouvernance de qualité**

La gestion des organisations et entités publiques (mais aussi privées) subissent des pressions constantes exercées par des lobbyistes internes et externes. Dans ce contexte, en vue de rationaliser la mise en œuvre des réformes structurelles, d'améliorer l'efficacité, de favoriser l'innovation ou d'atteindre des objectifs exigeants, la capacité des cadres supérieurs à (i) Exercer un leadership efficace pendant les périodes de transformation et de changement et à (ii) comprendre les règles de gestion, la culture (ou les us et coutumes) et la dynamique des organisations pourrait faire la différence entre l'échec organisationnel et le succès. Ainsi parle-t-on du binôme leadership-gouvernance.

*Dans ce cas spécifique, comment définir les notions de leadership et de gouvernance?*

La littérature sur le leadership, le management et la gouvernance propose un ensemble quasiment innombrable, mais aussi complémentaire des définitions sur le leadership et la gouvernance. Celles proposées dans Matata et Tsasa (2019) tiennent compte de la commodité analytique, au regard notamment des théorèmes 1 et 2 énoncés dans ledit papier[59]. Au point (F.1), nous présentons brièvement ces définitions. Au point (F.2), nous proposons une version explicite de ces définitions dites « analytiquement commodes ». Finalement, au point (F.3), nous présentons la définition classique du leadership et de la gouvernance.

**F.1. Définitions analytiquement commodes**

Leadership: volonté et capacité pour un décideur politique de résister ou de négocier stratégiquement avec les groupes de pression avec comme objectif majeur: prioriser l'atteinte des objectifs favorables au bien-être collectif.

Gouvernance: volonté et capacité pour un décideur politique de conduire avec succès des réformes structurelles favorables au bien-être collectif.

Dans ce cas, lorsque la volonté et la capacité sont effectives, c'est-à-dire se traduisent par des résultats concrets, alors on parle de leadership éclairé et de la gouvernance de qualité.

**F.2. Définitions analytiquement explicites**

Plus explicitement le leadership éclairé (ou efficace) peut être défini comme la capacité d'un dirigeant politique (i) d'avoir une vision claire du progrès de son pays, partagée par l'ensemble ou la majeure partie de la population, (ii) de définir des objectifs globaux et intermédiaires précis et les mettre en œuvre, (iii) de résister aux pesanteurs de toute nature pouvant fragiliser ou anéantir la démarche entreprise et (iv) de vaincre les obstacles érigés par les groupes de pression et autres lobbyistes et rentiers qui cherchent à maintenir le pays dans une trappe compatible avec leurs intérêts privés; alors que l'objectif ultime pour un leader éclairé est d'améliorer de manière durable le bien-être collectif, i.e. de l'ensemble de la population.

---

[59] En effet, à l'aide d'un modèle d'interactions stratégiques inspiré des jeux bayésiens, il a été démontré dans ce papier qu'une amélioration du binôme leadership-gouvernance peut aider à discipliner le comportement des groupes d'intérêts (théorème 1) et à générer une amélioration de Pareto (théorème 2) dans la gestion des ressources naturelles. Ce résultat peut, naturellement, être généralisé dans le cas de la gestion macroéconomique.

En parallèle, une gouvernance de qualité (ou simplement la bonne gouvernance) peut être comprise comme la capacité du gouvernement à gérer efficacement ses ressources, à mettre en œuvre des politiques pertinentes et à promouvoir le respect des institutions par les citoyens et l'État, ainsi que l'existence d'un contrôle démocratique sur ceux qui gouvernent.

### F.3. Définitions classiques

De manière classique, le leadership suppose la prise des initiatives conformément à une vision, celle du leader, mais aussi suppose la mobilisation d'un sous-ensemble d'agents œuvrant harmonieusement en vue d'atteindre des objectifs spécifiques (cf. Zaleznik 1977, Maccoby 2000).

*In fine*, la gouvernance est généralement perçue comme la manière dont les sociétés, les gouvernements et les organisations sont gérés et dirigés (cf. Edwards 2012) ou comme les traditions et les institutions par lesquelles l'autorité d'un pays est exercée (cf. Kaufman et al. 1999). Du point de vue de l'économie politique, la gouvernance fait référence à des éléments essentiels du vaste groupe d'institutions (cf. Acemoglu et al. 2008, Fukuyama 2013).

### Annexe G. Description de la base de données EPIN de la Banque mondiale

La base de données des indices d'évaluation de la qualité des politiques et des institutions nationales (EPIN ou CPIA en anglais, *Country Policy And Institutional Assessment*) classe les pays en fonction d'un ensemble de critères regroupés en quatre blocs: (i) Gestion économique; (ii) Politiques structurelles; (iii) Politiques d'inclusion sociale et d'équité; et (iv) Gestion et les institutions du secteur public. Plus explicitement, la base de données EPIN publie le classement suivant une échelle de 1 à 6 des performances dans les domaines suivants: (1) Classement de la transparence, de la responsabilisation et de la corruption dans le secteur public; (2) Classement du commerce; (3) Classement de la moyenne collective des politiques structurelles; (4) Classement de la moyenne collective des politiques relatives à l'inclusion sociale/équité; (5) Classement de l'efficience de la mobilisation des revenus; (6) Classement de la moyenne collective de la gestion du secteur public et des institutions; (7) Classement de la protection sociale; (8) Classement des droits de propriétés et de règles de gouvernance; (9) Classement de l'équité de l'utilisation de ressources publiques; (10) Classement de la qualité de l'administration publique; (11) Classement de la gestion macroéconomique; (12) Indice d'allocation des ressources de l'IDA; (13) Classement du renforcement des ressources humaines; (14) Classement de la politique sur l'égalité des sexes; (15); Classement de la politique fiscale; (16) Classement du secteur financier; (17) Classement de la qualité de la gestion budgétaire et financière; (18) Classement des politiques et institutions pour la durabilité de l'environnement; (19) Classement de la moyenne collective de la gestion économique; (20) Classement de la politique sur la dette; (21) Classement de l'environnement de réglementation des activités commerciales.

Les critères sont axés sur l'équilibre entre la capture des facteurs clés qui favorisent la croissance et la réduction de la pauvreté, avec la nécessité d'éviter une charge excessive sur le processus d'évaluation. Cahque critère est donc fondé sur les politiques et les arrangements institutionnels soumis au contrôle des décideurs politiques. Pour chaque critère, les performances des pays sont notées sur une échelle allant de 1 (faible) à 6 (fort). Le personnel de la Banque mondiale calcule ces notes en se basant sur un certain nombre d'indicateurs, d'observations et de jugements.

Le tableau G.1 présente la base de données des indices pour la RDC pour la période 2005-2018. La dernière est la moyenne arithmétique de l'ensemble des indices pour chaque année. C'est cette moyenne qui a été utilisée pour générer la figure 6 dans le texte.

Tableau G.1: Indices EPIN en RDC entre 2005 et 2018

| | 2005 | 2006 | 2007 | 2008 | 2009 | 2010 | 2011 | 2012 | 2013 | 2014 | 2015 | 2016 | 2017 | 2018 |
|---|---|---|---|---|---|---|---|---|---|---|---|---|---|---|
| Corruption dans le secteur public | 2.00 | 2.00 | 2.00 | 2.00 | 2.00 | 2.00 | 2.00 | 2.00 | 2.00 | 2.00 | 2.00 | 2.00 | 2.00 | 2.00 |
| Commerce | 4.00 | 4.00 | 4.00 | 4.00 | 3.50 | 3.00 | 3.00 | 3.00 | 3.50 | 3.50 | 3.50 | 3.50 | 3.50 | 3.50 |
| Politiques structurelles | 3.00 | 3.00 | 3.00 | 2.70 | 2.50 | 2.33 | 2.50 | 2.67 | 3.00 | 3.00 | 3.00 | 3.00 | 3.00 | 3.00 |
| Inclusion sociale | 2.90 | 2.90 | 2.90 | 2.90 | 2.80 | 2.80 | 2.80 | 2.80 | 2.80 | 2.90 | 3.00 | 3.00 | 3.00 | 3.10 |
| Mobilisation des revenus | 2.50 | 2.50 | 2.50 | 2.50 | 2.50 | 2.50 | 2.50 | 2.50 | 2.50 | 3.00 | 3.00 | 3.00 | 3.00 | 3.00 |
| Gestion du secteur public | 2.30 | 2.30 | 2.30 | 2.20 | 2.20 | 2.20 | 2.20 | 2.20 | 2.40 | 2.50 | 2.50 | 2.50 | 2.50 | 2.50 |
| Protection sociale | 3.00 | 3.00 | 3.00 | 3.00 | 3.00 | 2.50 | 2.50 | 2.50 | 2.50 | 2.50 | 2.50 | 2.50 | 2.50 | 2.50 |
| Règles de gouvernance | 2.00 | 2.00 | 2.00 | 2.00 | 2.00 | 2.00 | 2.00 | 2.00 | 2.00 | 2.00 | 2.00 | 2.00 | 2.00 | 2.00 |
| Utilisation de ressources publiques | 3.00 | 3.00 | 3.00 | 3.00 | 3.00 | 3.00 | 3.00 | 3.00 | 3.00 | 3.50 | 3.50 | 3.50 | 3.50 | 3.50 |
| Qualité de l'administration publique | 2.50 | 2.50 | 2.50 | 2.00 | 2.00 | 2.00 | 2.00 | 2.00 | 2.50 | 2.50 | 2.50 | 2.50 | 2.50 | 3.00 |
| Gestion macroéconomique | 3.50 | 3.50 | 3.50 | 3.50 | 3.50 | 3.50 | 3.50 | 3.50 | 3.50 | 3.50 | 3.50 | 3.00 | 2.50 | 3.00 |
| Allocation des ressources de l'IDA | 2.84 | 2.84 | 2.84 | 2.70 | 2.67 | 2.67 | 2.67 | 2.71 | 2.88 | 2.98 | 3.00 | 2.92 | 2.83 | 2.94 |
| Renforcement ressources humaines | 3.00 | 3.00 | 3.00 | 3.00 | 3.00 | 3.50 | 3.50 | 3.50 | 3.50 | 3.50 | 3.50 | 3.50 | 3.50 | 3.50 |
| Politique sur l'égalité des sexes | 3.00 | 3.00 | 3.00 | 3.00 | 2.50 | 2.50 | 2.50 | 2.50 | 2.50 | 2.50 | 2.50 | 2.50 | 2.50 | 3.00 |
| Politique fiscale | 3.50 | 3.50 | 3.50 | 3.50 | 3.50 | 3.50 | 3.50 | 3.50 | 3.50 | 3.50 | 3.50 | 3.00 | 2.50 | 3.00 |
| Secteur financier | 2.00 | 2.00 | 2.00 | 2.00 | 2.00 | 2.00 | 2.00 | 2.50 | 2.50 | 2.50 | 2.50 | 2.50 | 2.50 | 2.50 |
| Qualité de la gestion budgétaire | 2.50 | 2.50 | 2.50 | 2.50 | 2.50 | 2.50 | 2.50 | 2.50 | 3.00 | 3.00 | 3.00 | 3.00 | 3.00 | 2.50 |
| Durabilité de l'environnement | 2.50 | 2.50 | 2.50 | 2.50 | 2.50 | 2.50 | 2.50 | 2.50 | 2.50 | 3.00 | 3.00 | 3.00 | 3.00 | 3.00 |
| Gestion économique | 3.17 | 3.17 | 3.17 | 3.20 | 3.17 | 3.33 | 3.17 | 3.17 | 3.33 | 3.50 | 3.50 | 3.17 | 2.83 | 3.17 |
| Politique sur la dette | 2.50 | 2.50 | 2.50 | 2.50 | 2.50 | 3.00 | 2.50 | 2.50 | 3.00 | 3.50 | 3.50 | 3.50 | 3.50 | 3.50 |
| Réglementation des activités commerciales | 3.00 | 3.00 | 3.00 | 2.00 | 2.00 | 2.00 | 2.50 | 2.50 | 3.00 | 3.00 | 3.00 | 3.00 | 3.00 | 3.00 |
| Moyenne de l'indice EPIN | 2.80 | 2.80 | 2.80 | 2.70 | 2.63 | 2.63 | 2.63 | 2.67 | 2.83 | 2.92 | 2.95 | 2.88 | 2.82 | 2.91 |

Note : Pour chaque indice, les performances d'un pays sont notées sur une échelle allant de 1 (performances faibles ou médiocres) à 6 (performances élevées). Source : Banque mondiale (WDI DataBank 2019).

*In fine*, il sied de noter que les notes que reflètent les indices EPIN jouent un rôle essentiel dans la répartition, basée sur les performances, des ressources de l'IDA par la Banque mondiale, qui visent à aider les pays les plus pauvres du monde à stimuler la croissance et à promouvoir une prospérité partagée.

Fin des annexes.